\documentclass[sigconf]{acmart}

\usepackage[utf8]{inputenc} 
\usepackage[T1]{fontenc}    
\usepackage{hyperref}       
\usepackage{url}    
\usepackage{xspace}
\usepackage{subfiles}
\usepackage{booktabs}       
\usepackage{amsfonts}       
\usepackage{nicefrac}       
\usepackage{microtype}      
\usepackage{xcolor}         
\usepackage{graphicx}
\usepackage{amsmath}
\usepackage{subcaption}
\usepackage{multirow}
\usepackage{wrapfig}
\usepackage{sidecap}
\usepackage{lipsum}
\usepackage{enumitem}   

\newcommand{\modeltitle}{\textsc{DisProtBench}}
\newcommand{\model}{\text{DisProtBench}\xspace}
\newcommand{\res}[2]{#1 {\tiny$\pm$ #2}}
\usepackage{pifont}

\AtBeginDocument{%
  }

\begin{document}

\title{\modeltitle: Uncovering the Functional Limits of Protein Structure Prediction Models in Intrinsically Disordered Regions}

\author{
 \textbf{Xinyue Zeng\textsuperscript{1,*}},
 \textbf{Tuo Wang\textsuperscript{1,*}},
 \textbf{Adithya Kulkarni\textsuperscript{2}},
 \textbf{Alexander Lu\textsuperscript{1}},
 \textbf{Alexandra Ni\textsuperscript{1}},
 \textbf{Phoebe Xing\textsuperscript{1}},
 \textbf{Junhan Zhao\textsuperscript{3}},
 \textbf{Siwei Chen\textsuperscript{4}},
 \textbf{Dawei Zhou\textsuperscript{1}}
 \\
 \textsuperscript{1} Department of Computer Science, Virginia Polytechnic Institute and State University \\
 \textsuperscript{2} Department of Computer Science, Ball State University \\
 \textsuperscript{3} Department of Pediatrics, University of Chicago; Comprehensive Cancer Center, University of Chicago Medicine; Department of Biomedical Informatics, Harvard Medical School \\
 \textsuperscript{4} Department of Human Genetics, University of Chicago; Stanley Center for Psychiatric Research, Broad Institute of MIT and Harvard
}

\renewcommand{\shortauthors}{Zeng et al.}


\begin{abstract}
Intrinsically disordered regions (IDRs) play central roles in cellular function, yet remain poorly evaluated by existing protein structure prediction benchmarks. Current evaluations largely focus on well-folded domains, overlooking three fundamental challenges in realistic biological settings: the structural complexity of proteins, the resulting low availability of reliable ground truth,
and prediction uncertainty that can propagate into high-risk downstream failures, such as in drug discovery, protein-protein interaction modeling, and functional annotation.
We present \model, an IDR-centric benchmark that explicitly incorporates prediction uncertainty into the evaluation of protein structure prediction models (PSPMs). To address structural complexity and ground-truth scarcity, we curate and unify a large-scale, multi-modal dataset spanning disease-relevant IDRs, GPCR-ligand interactions, and multimeric protein complexes. To assess predictive uncertainty, we introduce \textit{Functional Uncertainty Sensitivity (FUS)}, a novel prediction uncertainty-stratified metric that quantifies downstream task performance under prediction uncertainty.
Using this benchmark, we conduct a systematic evaluation of state-of-the-art PSPMs and reveal clear, task-dependent failure modes. Protein-protein interaction prediction degrades sharply in IDRs, while structure-based drug discovery remains comparatively robust. These effects are largely invisible to standard global accuracy metrics, which overestimate functional reliability under prediction uncertainty.
We have open-sourced our benchmark and the codebase at \url{https://github.com/Susan571/DisProtBench}.
\end{abstract}




\thanks{* Both authors contributed equally to this research.}
\maketitle
\section{Introduction}
\label{sec:introduction}
Many critical cellular processes-including signal transduction, transcriptional regulation, and molecular recognition-are mediated not by static, well-folded protein domains, but by intrinsically disordered regions (IDRs)~\cite{bondos2022intrinsically, trivedi2022intrinsically, madhurima2023functional}. Unlike structured domains, IDRs lack stable conformations, exhibit high conformational variability, and engage in context-dependent interactions~\cite{wright2015intrinsically, uversky2019intrinsically}. These properties make IDRs indispensable for biological function, particularly in protein-protein interactions and dynamic assemblies, yet they remain challenging to characterize experimentally and computationally.
Recent advances in deep learning have led to substantial progress in protein structure prediction models (PSPMs). Systems such as AlphaFold (AF)~\cite{jumper2021highly, hu2024spatialppi} and ESMFold~\cite{lin2023evolutionary} achieve near-atomic accuracy on well-ordered protein domains and have been widely adopted in structure-based drug discovery and protein function annotation~\cite{chang2024revolutionizing, anfinsen1973principles, dill2008protein}. These successes have accelerated biological discovery and expanded the practical use of predicted structures in downstream applications. 

\begin{figure*}[h!]
    \centering
    \includegraphics[width=0.9\linewidth]{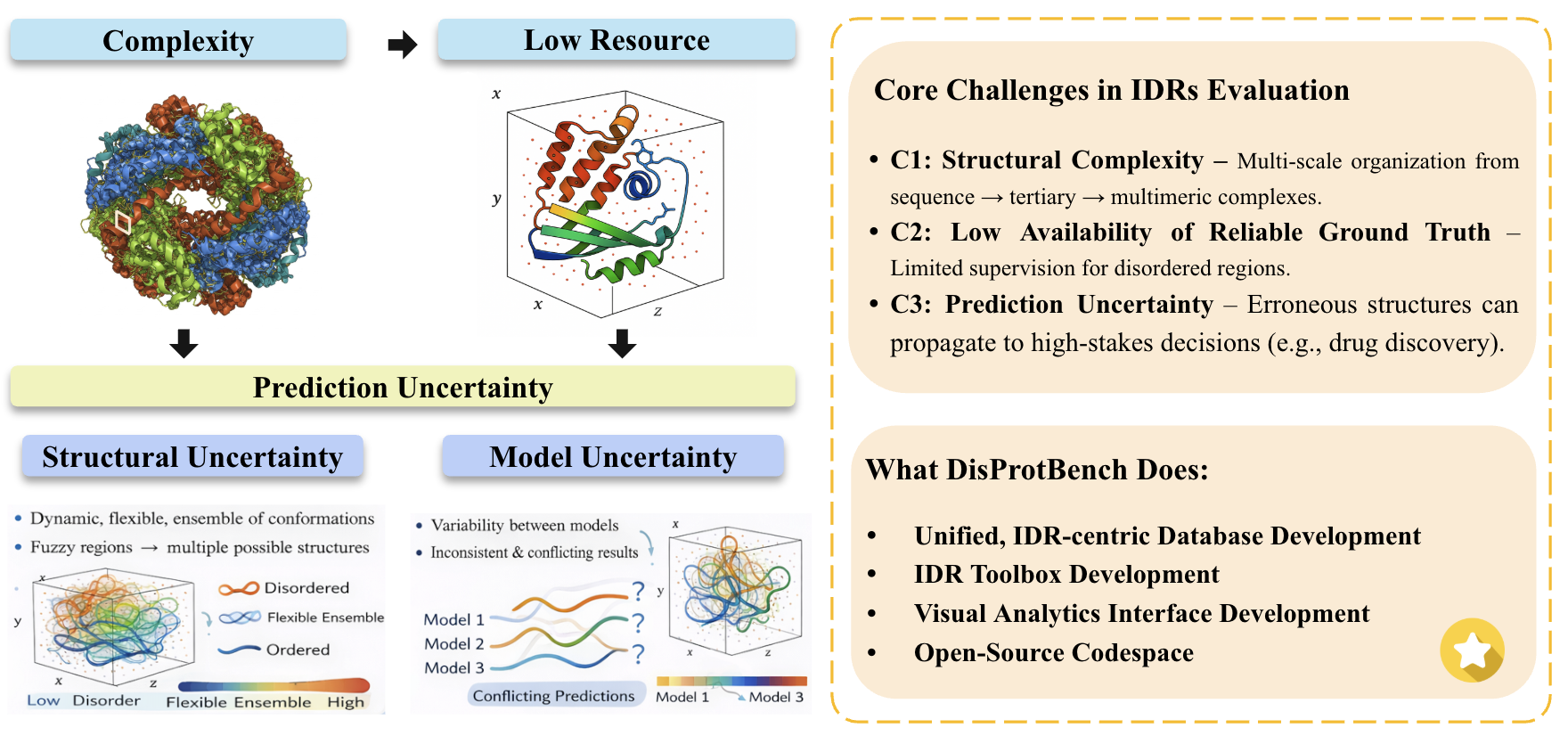}
    \caption{Overview of Intrinsically Disordered Regions (IDRs). Structural complexity and low availability of reliable ground truth in IDRs induce prediction uncertainty that propagates to downstream tasks, motivating IDR-centric functional evaluation.}
    \label{fig:background}
    \vspace{-1em}
\end{figure*}



Despite these advances, existing evaluation benchmarks fail to provide diagnostic visibility into IDR-centric regimes. Standard frameworks such as the Critical Assessment of Structure Prediction (CASP) primarily assess global folding accuracy using aggregate structural metrics~\cite{moult2018critical}. However, recent studies have shown that such metrics can obscure substantial model uncertainty and result in task-relevant failures in IDRs, which are induced by their structural complexity and low availability of reliable ground truth~\cite{krokidis2025alphafold3, piovesan2022intrinsic}. As a result, models may appear globally accurate while failing to capture the flexible conformational behavior that underpins downstream biological functions.
As illustrated in Figure~\ref{fig:background}, structural uncertainty introduced at the residue or domain level can propagate to interaction interfaces and multimeric assemblies, where IDR-induced variability directly undermines task-level evaluation. This limitation is particularly consequential for applications that depend on dynamic and context-dependent interactions, including protein-protein interaction (PPI) prediction and phase separation~\cite{zhang2025protein}. Together, these observations indicate that current benchmarks are insufficient for assessing the functional reliability of PSPMs in the presence of IDRs.

Together, these limitations expose three tightly coupled challenges in evaluating PSPM: 
\textbf{\textit{(i) Structural Complexity:}} Proteins enriched with IDRs exhibit high structural complexity, characterized by heterogeneous conformational ensembles, transient interfaces, and strong context dependence, which violates the static structure assumptions underlying most existing evaluation frameworks;
\textbf{\textit{(ii) Low Availability of Reliable Ground Truth:}} This complexity directly leads to a scarcity of reliable ground-truth structures: IDR-centric regions are underrepresented in experimentally resolved datasets, rendering structure-level supervision and reference-based evaluation unreliable or infeasible at scale;
\textbf{\textit{(iii) Prediction Uncertainty:}} Under low availability of reliable ground truth, PSPMs inevitably operate under substantial predictive uncertainty, which, if left unexamined, can propagate into downstream biomedical applications, such as protein-protein interaction analysis and drug discovery, where erroneous structural predictions may induce high-risk functional or clinical failures.

To address the challenges, we introduce \model, an IDR-centric benchmark and evaluation framework for protein structure prediction. Rather than optimizing raw accuracy, \model diagnoses functional reliability under structural uncertainty through unified data curation, uncertainty-aware evaluation, and interactive diagnostics. Our contributions are summarized as follows:

\begin{itemize}[leftmargin=9pt]
\item \textbf{Database Development.} 
    We collect and unify a large-scale, multi-modal dataset spanning clinically relevant intrinsically disordered regions (IDRs), GPCR-ligand interaction pairs, and multimeric protein complexes. 
    This unified representation explicitly captures structural complexity while mitigating ground-truth scarcity by enabling task-level evaluation without assuming a single native structure.
\item \textbf{IDR Toolbox Development.} 
    We propose a diagnostic metric that quantifies downstream task performance under prediction uncertainty. 
    This metric enables principled assessment of predictive risk induced by IDRs, beyond what is revealed by aggregate accuracy metrics.
\item \textbf{Visual Analytics Interface Development.} 
    We provide an integrated evaluation toolbox and web-based portal that support reproducible benchmarking, fine-grained error analysis, and human-in-the-loop exploration, allowing users to trace prediction uncertainty to downstream task failures.
\item \textbf{Open-Sourced Codebase.} 
    For accessibility and reproducibility, we have open-sourced our benchmark and the codebase at \url{https://github.com/Susan571/DisProtBench}.
\end{itemize}

\section{Preliminary}
\label{sec:prelim}
\subsection{Protein Structure Prediction Models}
\begin{figure}[h!]
    \centering
    \includegraphics[width=0.75\linewidth]{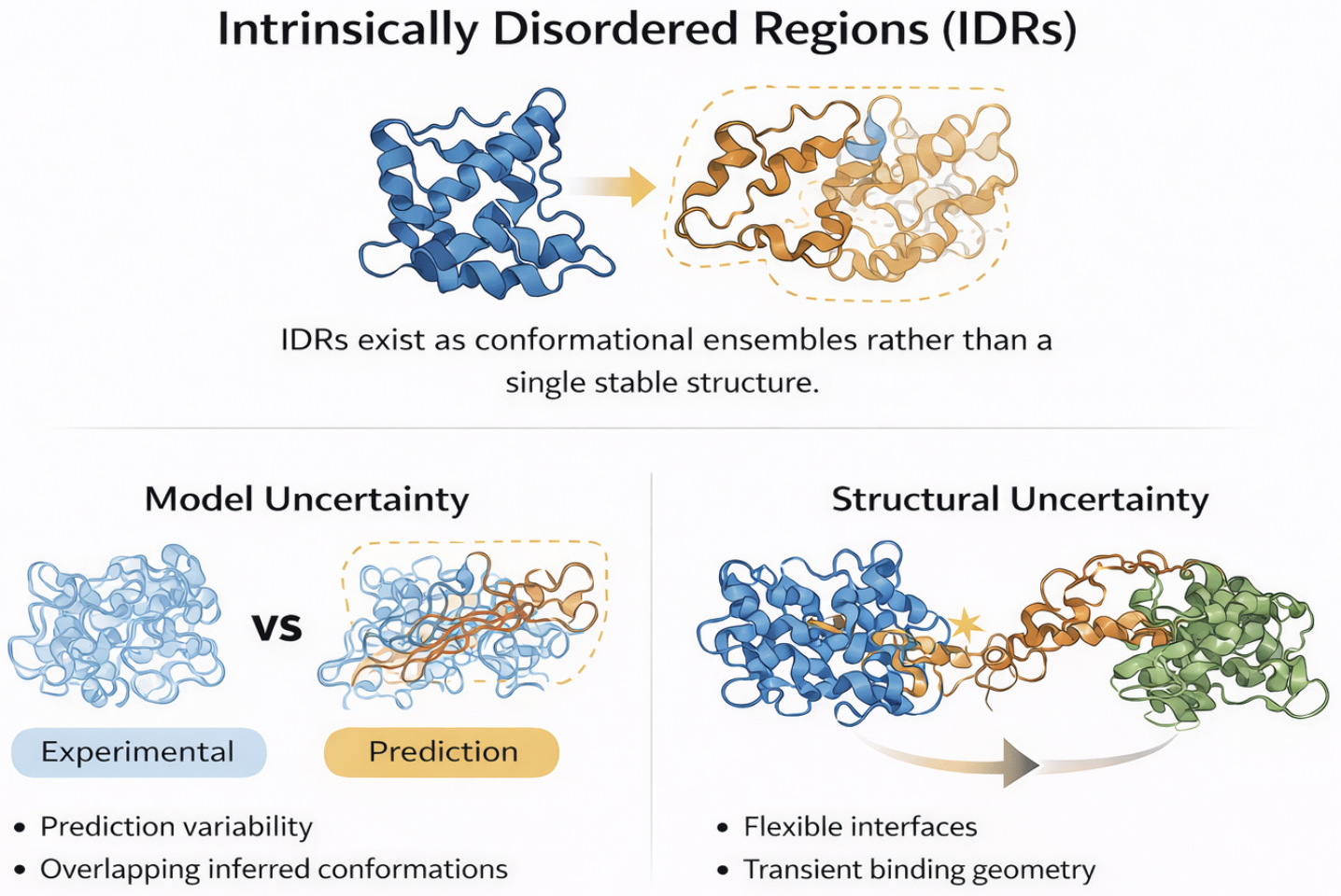}
    \caption{
    Overview of intrinsically disordered regions (IDRs).
    }
    \label{fig:idr_illustration}
    \vspace{-1em}
\end{figure}

\begin{figure*}[h!] 
\centering 
\includegraphics[width=0.85\linewidth]{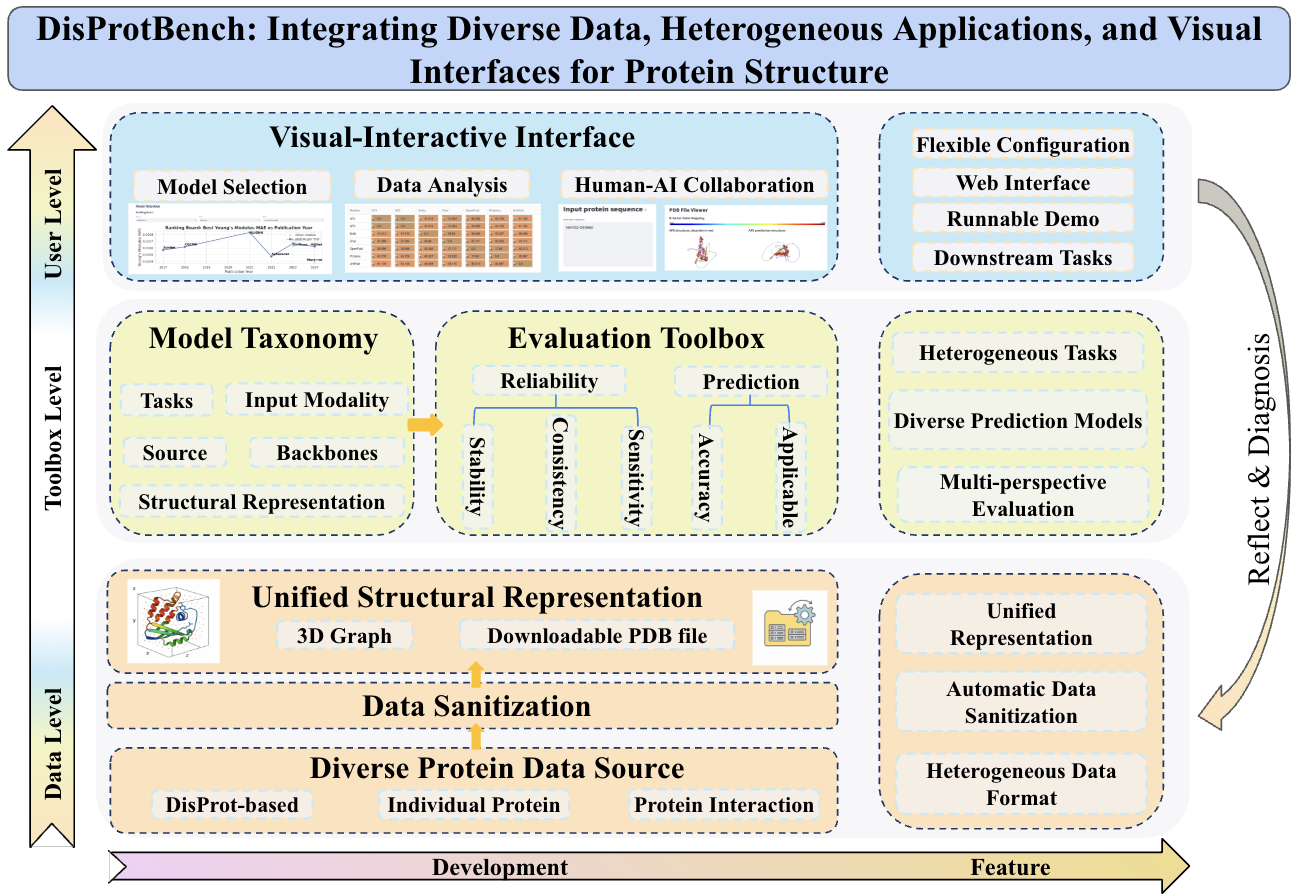} 
\caption{ Overview of \model. We structured the benchmark across three levels: Input Complexity (Data), Functional Utility (Task), and Interpretability (User), allowing us to isolate specific sources of error in disordered regions.} 
\label{fig:overview} 
\vspace{-1em}
\end{figure*}
Protein structure prediction models (PSPMs) infer three-dimensional protein structures directly from amino acid sequences. The field has progressed from template-based methods such as Rosetta~\cite{rohl2004protein} to end-to-end deep learning systems trained on large structural corpora. Early evaluation frameworks, formalized through CASP~\cite{moult1995large, moult2018critical}, introduced metrics such as RMSD and GDT-TS that implicitly assume a single, well-defined native structure per sequence.
Recent deep learning PSPMs, led by AlphaFold2~\cite{jumper2021highly,abramson2024alphafold3} and RoseTTAFold~\cite{baek2021accurate}, achieve near-atomic accuracy on well-folded domains, with later models such as ESMFold~\cite{lin2023evolutionary} and AlphaFold3~\cite{abramson2024alphafold3} extending predictions to complexes and protein-ligand interactions. These advances have driven widespread adoption in downstream applications, including drug discovery and function annotation.
However, both training and evaluation remain dominated by crystallographic, rigid domains, biasing assessment toward static conformations. Consequently, the reliability of PSPMs in structurally complex and disorder-rich regions is poorly captured by existing evaluation frameworks.


\subsection{Intrinsically Disordered Regions}
Intrinsically disordered regions (IDRs) are protein segments that do not adopt a single stable three-dimensional structure, but instead populate context-dependent conformational ensembles~\cite{wright2015intrinsically, uversky2019intrinsically, oldfield2014intrinsically}. As illustrated in Figure~\ref{fig:idr_illustration}, IDRs coexist with structured domains yet exhibit flexible, multi-state conformations that enable dynamic protein-protein interactions and context-dependent functional roles.
This ensemble behavior fundamentally departs from the classical structure-function paradigm that assumes a unique native fold. In IDR-centric regions, deviations from a single reference structure-treated as error by standard metrics-are expected and often biologically meaningful, reflecting overlapping conformations and transient interaction geometries rather than prediction failure. Recent studies confirm that conventional structural accuracy metrics are ill-suited to capture this heterogeneity~\cite{lotthammer2024direct, erdHos2024deep}, making structural discrepancy a weak proxy for functional reliability in IDRs. These limitations motivate evaluation frameworks that explicitly account for disorder-induced uncertainty, forming the basis of the disorder-aware benchmarks and diagnostics introduced in this work.

\subsection{Prediction Uncertainty}
\textbf{Structural Uncertainty.}
Intrinsically disordered regions introduce structural uncertainty because they do not admit a single stable native conformation.
In such regions, multiple conformations may be biologically valid, and deviations from a single reference structure do not necessarily indicate functional error~\cite{vangone2015contacts, dhakal2022artificial, zhang2024machine}.
As a result, conventional structure-centric accuracy metrics are ill-suited for evaluating predictions in IDR settings.

\noindent \textbf{Model Uncertainty.}
Beyond biological variability, protein structure prediction models exhibit model uncertainty due to limited training data, inductive bias, and architectural constraints.
Many PSPMs provide internal confidence signals, such as AlphaFold’s pLDDT~\cite{jumper2021highly,evans2021protein, hu2024spatialppi}, which correlate with disorder but are not ground-truth annotations.
In this work, we treat model uncertainty as a diagnostic signal for uncertainty-aware evaluation, rather than as a proxy for true structural correctness.

\section{\model Overview}
\label{sec:overview}
Motivated by the limitations discussed in Section~\ref{sec:prelim}, \model is an IDR-centric evaluation benchmark for PSPMs, designed to assess functional reliability under structural uncertainty rather than structural accuracy alone.
As illustrated in Figure~\ref{fig:overview}, \model integrates heterogeneous protein data, downstream functional tasks, and diagnostic evaluation tools into a unified framework reflecting realistic biological complexity.
Its design is guided by three core principles, each addressing a fundamental challenge in evaluating PSPMs in IDR settings:

\begin{table*}[h!]
\centering
\small
\setlength{\tabcolsep}{6pt}
\caption{
Datasets in \model and their targeted structural uncertainty regimes.
Each dataset probes a distinct IDR-driven failure mode of PSPMs under task-specific evaluation.
}
\begin{tabular}{l|ccc|c}
\toprule
\textbf{Dataset} &
\textbf{Data Focus} &
\textbf{Uncertainty Source} &
\textbf{Eval Task} &
\textbf{Scale} \\
\midrule

DisProt-Based Dataset
& Long IDRs + Disease Variants
& Conformational Ensembles
& PPI
& $\sim$10$^3$ proteins / $\sim$10$^4$ PPIs \\

Individual Protein Dataset
& Ordered Pockets + IDRs
& Local vs. Global Uncertainty
& Drug Discovery
& $\sim$10$^5$ interactions \\

Protein Interaction Dataset
& Flexible Interfaces
& Transient Interface Geometry
& PPI
& $\sim$10$^3$ complexes \\

\bottomrule
\end{tabular}
\label{tab:dataset_overview}
\vspace{-1em}
\end{table*}

\begin{itemize}[leftmargin=9pt]
\item \textbf{Database Development.} 
Protein structures in real biological systems are highly heterogeneous, often involving IDRs that lack a single stable conformation.
\model explicitly targets proteins and interaction scenarios where disorder plays a functional role, including IDRs proteins, disorder-mediated interfaces, and multimeric complexes.
By moving beyond rigid, crystallography-dominated regimes, the benchmark captures structural complexity that is largely absent from existing evaluation datasets.
\item \textbf{IDR Toolbox development.}
In IDRs regions, definitive structural ground truth is often unavailable or ill-defined, making structure-only evaluation insufficient.
Instead of evaluating PSPMs in isolation, \model assesses predicted structures through their impact on downstream functional tasks, including protein-protein interaction (PPI) prediction, structure-based drug discovery, and multimeric complex analysis.
These tasks provide task-level supervision that serves as a practical proxy for functional correctness in the absence of reliable structural labels.
\item \textbf{Visual Analytics Interface Development}
To support interpretability in IDRs where aggregate metrics are insufficient, \model includes a visual analytics interface that links structural predictions, uncertainty estimates, and downstream task outcomes. The interface enables residue-level inspection with uncertainty overlays, side-by-side comparison across PSPMs, and direct attribution of task-level failures to localized structural uncertainty. This transforms the benchmark from a static evaluation suite into a diagnostic tool, enabling principled error analysis under structural ambiguity.
\end{itemize}
Together, these principles position \model as a benchmark for functional reliability under uncertainty in IDR settings, rather than as a conventional structure prediction contest.

\section{Database Development}
\label{sec:data}
\model integrates diverse protein data sources spanning multiple biological contexts (Figure~\ref{fig:overview}, bottom), with the explicit goal of capturing structural complexity and mitigating ground-truth scarcity in IDRs.
Rather than assuming a single canonical structure, the benchmark supports functional evaluation under heterogeneous and uncertain structural conditions.
It consists of three dataset categories, each targeting a distinct aspect of IDR-driven complexity:





\begin{itemize}[leftmargin=9pt]
    \item \textbf{DisProt-Based Dataset: IDRs in Human Disease.}
    We construct an IDR-enriched dataset by curating human proteins with long disordered segments ($\geq$20 residues) from UniProt, based on experimentally annotated IDRs regions.
    To introduce functional context in the absence of definitive structural ground truth, we integrate protein-protein interactions from HINT and pathogenic missense variants from ClinVar.
    Protein pairs are retained if both genes contain at least two disease-associated missense mutations.
    All sequences are normalized to canonical UniProt references and deduplicated.
    This dataset probes PSPM behavior in clinically relevant IDR settings rather than static fold accuracy.
    \item \textbf{Individual Protein Dataset: IDRs and Ligand Binding.}
    To assess reliability in structure-based drug discovery, we curate a GPCR-ligand interaction dataset from ChEMBL and BindingDB, including 71,757 interactions across the top 20 GPCR targets and 33,212 pain-related protein-ligand pairs, filtered by binding affinity ($pK_i \in [6,9]$).
    Ligands are standardized using InChIKey, proteins are matched by UniProt ID, and interactions are annotated by pharmacological role and compound type~\cite{yang2024deep}.
    This dataset reflects realistic drug discovery scenarios where ordered binding pockets coexist with IDR regions.
    \item \textbf{Protein Interaction Dataset: IDR-Mediated Interfaces.}
    To evaluate multimeric complexes with flexible interfaces, we construct a PPI dataset.
    Complexes are filtered by structural quality (median pLDDT $\geq 70$, pDockQ, interface compactness) and retained only when predicted interfaces overlap with annotated IDRs.
    All complexes are standardized into a unified representation for consistent interface definition and evaluation~\cite{hu2024spatialppi}.
\end{itemize}

\begin{table*}[h!]
\centering
\small
\setlength{\tabcolsep}{6pt}
\caption{
Models Toolbox. PSPMs evaluated in \model span diverse architectures, input modalities, and structural representations, enabling systematic analysis of robustness and failure modes under structural uncertainty.
}
\begin{tabular}{l|ccccc}
\toprule
\textbf{PSPM} &
\textbf{Supported Tasks} &
\textbf{Architecture} &
\textbf{Input Modality} &
\textbf{Structural Rep.} &
\textbf{Reference} \\
\midrule
AlphaFold2 (AF2)      & PPI, Drug & Evoformer              & MSA            & Atomic                 & \cite{jumper2021highly} \\
AlphaFold3 (AF3)      & PPI, Drug & Evoformer + LLM        & MSA + Sequence & Atomic + Ligand        & \cite{abramson2024alphafold3} \\
OpenFold               & PPI, Drug & Evoformer              & MSA            & Atomic                 & \cite{Ahdritz2024OpenFold} \\
UniFold                & PPI, Drug & Evoformer              & MSA            & Atomic                 & \cite{Li2022UniFold} \\
Boltz                  & PPI, Drug & Transformer            & Sequence-only  & Coarse-grained         & \cite{wohlwend2024boltz1} \\
Chai                   & PPI, Drug & Transformer            & Sequence-only  & Coarse-grained         & \cite{Chai-1-Technical-Report} \\
Proteinx               & PPI, Drug & Transformer+            & Sequence-only  & Atomic + Ligand        & \cite{chen2025protenix} \\
ESMFold                & Drug      & Transformer            & Sequence-only  & Coarse-grained         & \cite{lin2023evolutionary} \\
OmegaFold              & Drug      & Transformer            & Sequence-only  & Coarse-grained         & \cite{OmegaFold} \\
RoseTTAFold            & Drug      & Hybrid (CNN + Attn)    & MSA            & Atomic                 & \cite{baek2021accurate} \\
DeepFold               & Drug      & Custom DL              & Sequence-only  & Atomic                 & \cite{Lee2023DeepFold} \\
\bottomrule
\end{tabular}
\label{tab:models_toolbox}
\vspace{-1em}
\end{table*}

\begin{table}[t]
\centering
\small
\renewcommand{\arraystretch}{1.6}
\setlength{\tabcolsep}{7pt}
\caption{
Evaluation Toolbox. Task-level metrics across classification, regression, and structural interface prediction used to compute FUS, which quantifies downstream performance under confidence-stratified structural uncertainty.
}
\begin{tabular}{c | l | l}
\toprule
\textbf{Category} & \textbf{Metric} & \textbf{Definition} \\
\midrule
\multirow{4}{*}{Classification}
& Accuracy & $\frac{TP+TN}{TP+TN+FP+FN}$ \\
& Precision & $\frac{TP}{TP+FP}$ \\
& Recall & $\frac{TP}{TP+FN}$ \\
& F1 Score & $\frac{2TP}{2TP+FP+FN}$ \\
\midrule
\multirow{3}{*}{Regression}
& MAE & $\frac{1}{N}\sum_{i=1}^N |y_i-\hat{y}_i|$ \\
& MSE & $\frac{1}{N}\sum_{i=1}^N (y_i-\hat{y}_i)^2$ \\
& Pearson $R$ & $\mathrm{corr}(y,\hat{y})$ \\
\midrule
\multirow{4}{*}{Structural Interface}
& Receptor Precision (RP) &
$\frac{|\mathrm{Pred}_R \cap \mathrm{True}_R|}{|\mathrm{Pred}_R|}$ \\
& Receptor Recall (RR) &
$\frac{|\mathrm{Pred}_R \cap \mathrm{True}_R|}{|\mathrm{True}_R|}$ \\
& Ligand Precision (LP) &
$\frac{|\mathrm{Pred}_L \cap \mathrm{True}_L|}{|\mathrm{Pred}_L|}$ \\
& Ligand Recall (LR) &
$\frac{|\mathrm{Pred}_L \cap \mathrm{True}_L|}{|\mathrm{True}_L|}$ \\
\bottomrule
\end{tabular}
\label{tab:evaluation_toolbox}
\vspace{-1.5em}
\end{table}


\section{IDR Toolbox Development}
\label{sec:metrix}
In this section, we introduce an IDR-centric toolbox that operationalizes IDR-centric evaluation across models and metrics. Section~\ref{sec:model} presents the machine learning toolbox of PSPMs included in \model, while Section~\ref{sec:evaluation} details the evaluation toolbox, culminating in a novel IDR-centric diagnostic-Functional Uncertainty Sensitivity (FUS)-designed to capture functional risk induced by IDRs.

\subsection{Machine Learning Toolbox Development}
\label{sec:model}
To ensure broad coverage of contemporary protein structure prediction models (PSPMs), 
we benchmark a diverse set of state-of-the-art systems spanning heterogeneous architectures, 
input modalities, and structural representations, as summarized in Table~\ref{tab:models_toolbox}.
The selected PSPMs cover three major architectural paradigms:
(i) Evoformer-based models leveraging MSAs (e.g., AF2, AF3, OpenFold, UniFold),
(ii) sequence-only transformer models with coarse-grained representations (e.g., Boltz, Chai, ESMFold),
and (iii) hybrid or ligand-aware architectures that jointly model protein and small-molecule structures.
This diversity ensures that our benchmark probes not only performance variation across models,
but also architectural sensitivity to structural uncertainty.
\begin{figure*}[h!]
    \centering    \includegraphics[width=0.95\linewidth,height=15em]{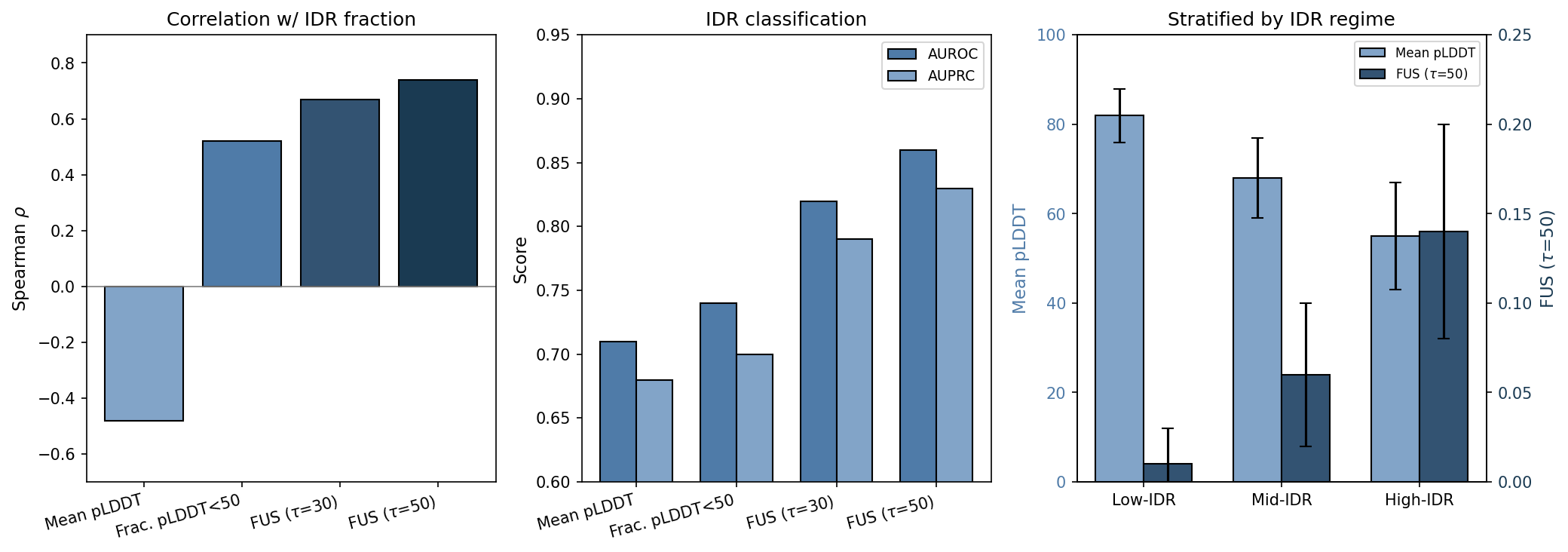}
    \caption{
    FUS vs. pLDDT on the DisProt-based dataset (Section~\ref{sec:data}). Left: Spearman correlation ($\rho$) between IDR fraction and uncertainty measures. Middle: IDR classification performance (AUROC, AUPRC). Right: IDR-stratified comparison of mean pLDDT (left axis) and Functional Uncertainty Sensitivity (FUS, $\tau=50$, right axis). FUS aligns more strongly with ground-truth IDRs and functional risk than confidence-based metrics (e.g., pLDDT), particularly for IDR-rich proteins.
    }
    \label{fig:case_fus_disprot}
    \vspace{-1em}
\end{figure*}

\subsection{Evaluation Toolbox Development}
\label{sec:evaluation}
\textbf{Existing Evaluation Methods.} 
Table~\ref{tab:evaluation_toolbox} summarizes the task-level metrics used throughout the benchmark.
Rather than introducing task-specific ad hoc scores, we adopt standard classification,
regression, and structural interface metrics that are widely used in downstream biological applications,
including PPI prediction and structure-based drug discovery.












\textbf{Functional Uncertainty Sensitivity (FUS).}
Current protein structure evaluation treats pLDDT as an exchangeable uncertainty signal, but in IDRs, uncertainty is structured and functionally localized, causing aggregate task metrics to mix signal with uncertainty-induced noise. To isolate this effect, we introduce \emph{Functional Uncertainty Sensitivity} (FUS), an IDR-centric diagnostic metric that measures how downstream task performance responds to uncertainty-driven structural perturbations. FUS is not an accuracy metric, but a sensitivity functional that captures the dependence of functional inference on structurally uncertain regions. 

Formally, let $\{(x_i, c_i)\}_{i=1}^n$ denote predicted structural features $x_i$ with associated uncertainty scores $c_i$, where $c_i$ corresponds to pLDDT or analogous model-provided uncertainty estimates. For a uncertainty threshold $\tau$, define the filtered structure
\[
x^{(\tau)} = \{x_i : c_i \ge \tau\},
\]
and let $\mathrm{Perf}_T(\tau) := \mathrm{Perf}_T(x^{(\tau)})$ denote a task-specific performance metric (e.g., accuracy/F1 for classification or $-\mathrm{MAE}$ for regression). We define
\begin{equation}
\mathrm{FUS}_T(\tau) = \mathrm{Perf}_T(\tau) - \mathrm{Perf}_T(0),
\end{equation}
where $\mathrm{Perf}_T(0)$ corresponds to evaluation on the unfiltered predicted structure. Equivalently, $\mathrm{FUS}_T(\tau)$ estimates the response of the functional $\mathrm{Perf}_T$ to a structured, uncertainty-ranked perturbation of its input, distinguishing uncertainty that is functionally aligned from uncertainty that is exchangeable.

\begin{figure*}[h!]
\centering
\includegraphics[width=\linewidth]{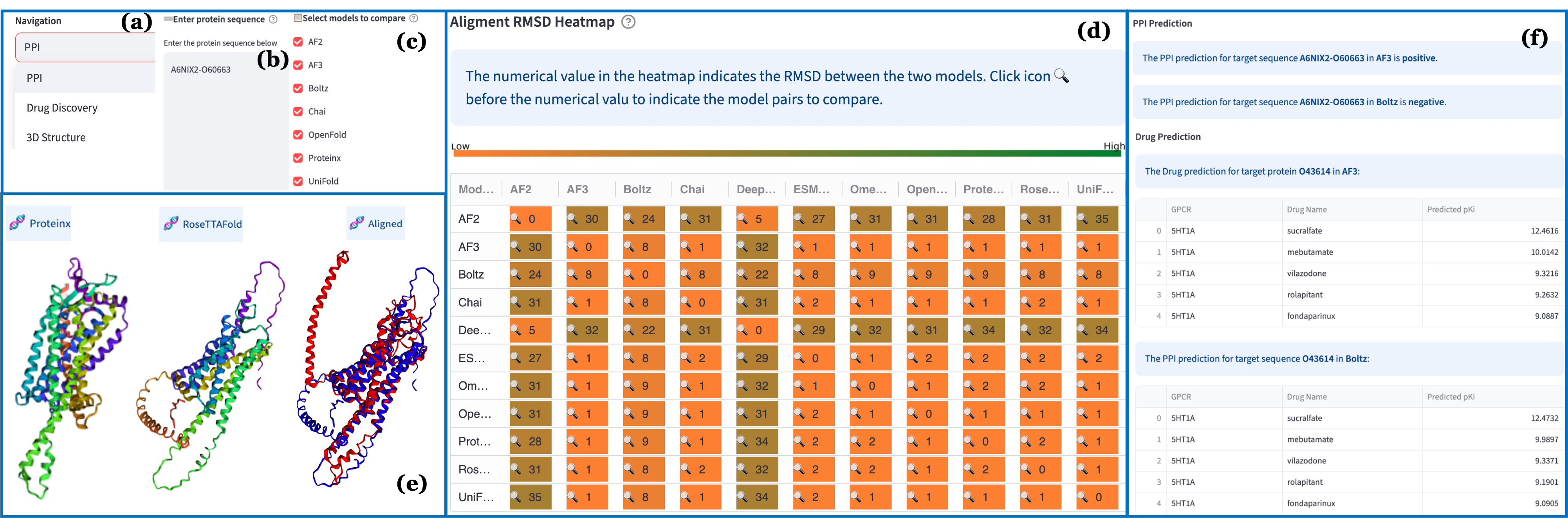}
\caption{Overview of the analytics portal. The interface supports (a-c) task and model selection, (d-e) uncertainty-aware structural diagnosis, and (f) downstream functional assessment, enabling direct linkage between structural uncertainty and task-level failure.}
\label{fig:UI}
\vspace{-1em}
\end{figure*} 

Statistically, $\mathrm{FUS}_T(\tau)$ measures the conditional performance shift induced by removing residues ranked as uncertainty score $c_i$, where scores such as pLDDT serve only as monotone stratification variables. Under this view, FUS approximates the directional derivative of $\mathrm{Perf}_T$ with respect to uncertainty-weighted structural perturbations. Within \model, FUS is used for diagnostic analysis rather than ranking, revealing IDR-driven failure modes among models with similar aggregate performance.

\textbf{Case Study of FUS.}
We conduct a focused case study on the DisProt-Based Dataset to compare pLDDT with our FUS against experimentally annotated IDRs. As shown in Figure~\ref{fig:case_fus_disprot}, while mean pLDDT exhibits a negative correlation with IDR fraction ($\rho=-0.48$), its ability to explain functional variability is limited: proteins with similar uncertainty profiles often display substantially different downstream behavior. Aggregating high-uncertainty residues improves alignment modestly (fraction pLDDT$<50$, $\rho=0.52$), but remains insufficient to capture functional risk. In contrast, FUS shows substantially stronger alignment with ground-truth IDRs, achieving $\rho=0.67$ at $\tau=30$ and $\rho=0.74$ at $\tau=50$, and significantly higher IDR classification performance (AUROC up to 0.86, AUPRC up to 0.83). Stratified analysis further reveals a clear functional separation: IDR-rich proteins exhibit sharply increased FUS despite overlapping pLDDT distributions, whereas low-IDR proteins remain functionally stable. These results demonstrate that FUS captures functional consequences of structural uncertainty rather than confidence alone, providing a more informative diagnostic for IDR-centric evaluation.

\section{Visual Interface Development}
\label{sec:portal}
\begin{figure*}[H]
\centering
\includegraphics[width=\linewidth]{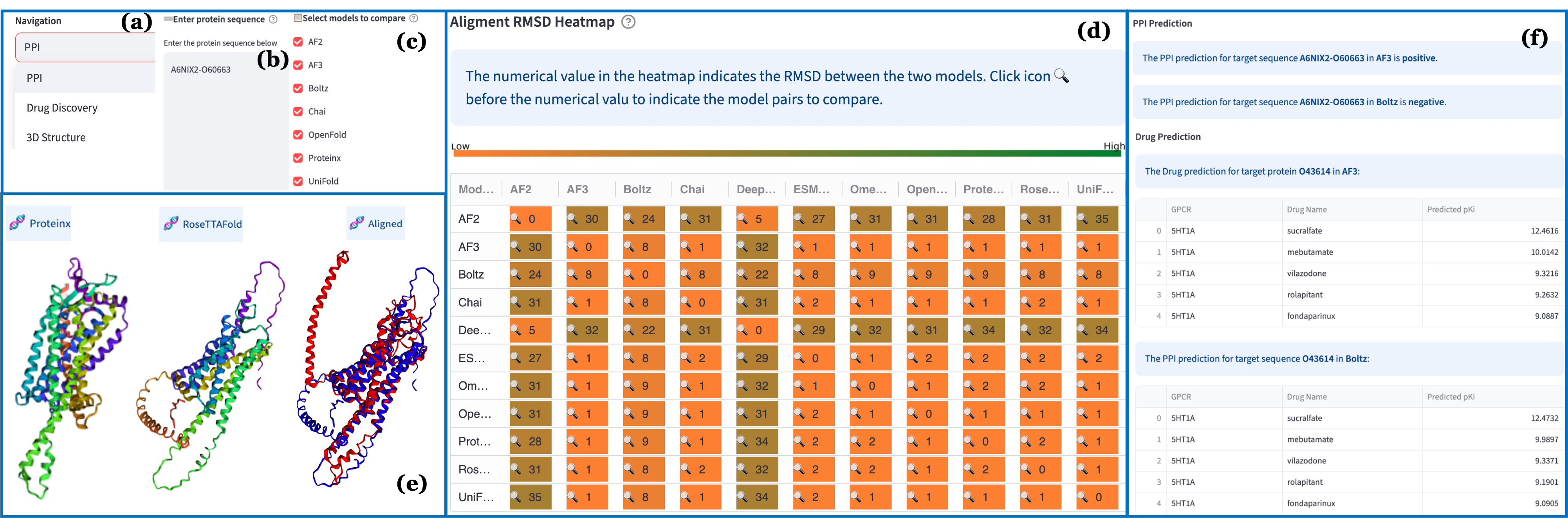}
\caption{Overview of the analytics portal. The interface supports (a--c) task and model selection, (d--e) confidence-aware structural diagnosis, and (f) downstream functional assessment, enabling direct linkage between structural uncertainty and task-level failure.}
\label{fig:UI}
\end{figure*}
To support reproducibility and community adoption, we provide an interactive portal (Figure~\ref{fig:UI}) as a companion to the \model benchmark. The portal operationalizes the IDR-centric evaluation protocol, rather than serving as an independent system contribution.
The interface exposes benchmark datasets, predicted structures, and evaluation outputs within a unified environment. Residue-level uncertainty visualization enables localization of structural uncertainty, while side-by-side comparison across PSPMs facilitates diagnosis of task-level errors arising from IDR-centric regions. Structural predictions are directly linked to downstream task outcomes, supporting function-aware analysis in both PPI and drug discovery settings.
We have open-sourced our benchmark and the codebase at \url{https://github.com/Susan571/DisProtBench}, enabling transparent reproduction and extension of our analyses without requiring local inference pipelines.


\begin{table*}[h!]
\centering
\small
\setlength{\tabcolsep}{4pt} 
\caption{Performance comparison on PPI prediction across PSPMs. FUS columns report performance shifts relative to the original prediction after confidence-ranked filtering ($\mathrm{FUS}_T(\tau)$).}
\begin{tabular}{l|cccc|cccc|cccc}
\toprule
\multirow{2}{*}{\textbf{PSPM}} &
\multicolumn{4}{c|}{\textbf{Original}} &
\multicolumn{4}{c|}{\textbf{FUS ($\tau=30$)}} &
\multicolumn{4}{c}{\textbf{FUS ($\tau=50$)}} \\
& \textbf{Acc} & \textbf{Prec} & \textbf{Rec} & \textbf{F1}
& \textbf{Acc} & \textbf{Prec} & \textbf{Rec} & \textbf{F1}
& \textbf{Acc} & \textbf{Prec} & \textbf{Rec} & \textbf{F1} \\
\midrule
AF2  & 0.793 & 0.783 & 0.799 & 0.791 & 0.802 & 0.791 & 0.812 & 0.801 & 0.818 & 0.809 & 0.825 & 0.817 \\
AF3       & 0.902 & 0.888 & 0.915 & 0.901 & 0.905 & 0.893 & 0.905 & 0.906 & 0.913 & 0.8989 & 0.93 & 0.914 \\
Boltz     & 0.850 & 0.848 & 0.853 & 0.850 & 0.858 & 0.853 & 0.863 & 0.858 & 0.869 & 0.870 & 0.868 & 0.869 \\
Chai      & 0.850 & 0.841 & 0.863 & 0.852 & 0.858 & 0.847 & 0.873 & 0.860 & 0.869 & 0.857 & 0.887 & 0.871 \\
OpenFold  & 0.624 & 0.605 & 0.605 & 0.605 & 0.643 & 0.622 & 0.638 & 0.630 & 0.671 & 0.656 & 0.651 & 0.653 \\
Proteinx  & 0.810 & 0.809 & 0.812 & 0.810 & 0.819 & 0.820 & 0.818 & 0.819 & 0.834 & 0.834 & 0.835 & 0.834 \\
UniFold   & 0.552 & 0.378 & 0.667 & 0.483 & 0.567 & 0.389 & 0.667 & 0.491 & 0.597 & 0.417 & 0.714 & 0.526 \\
\bottomrule
\end{tabular}
\label{tab:combined_ppi_metrics}
\vspace{-1em}
\end{table*}

\begin{figure*}[h!]
    \centering   
    \includegraphics[width=0.9\linewidth]{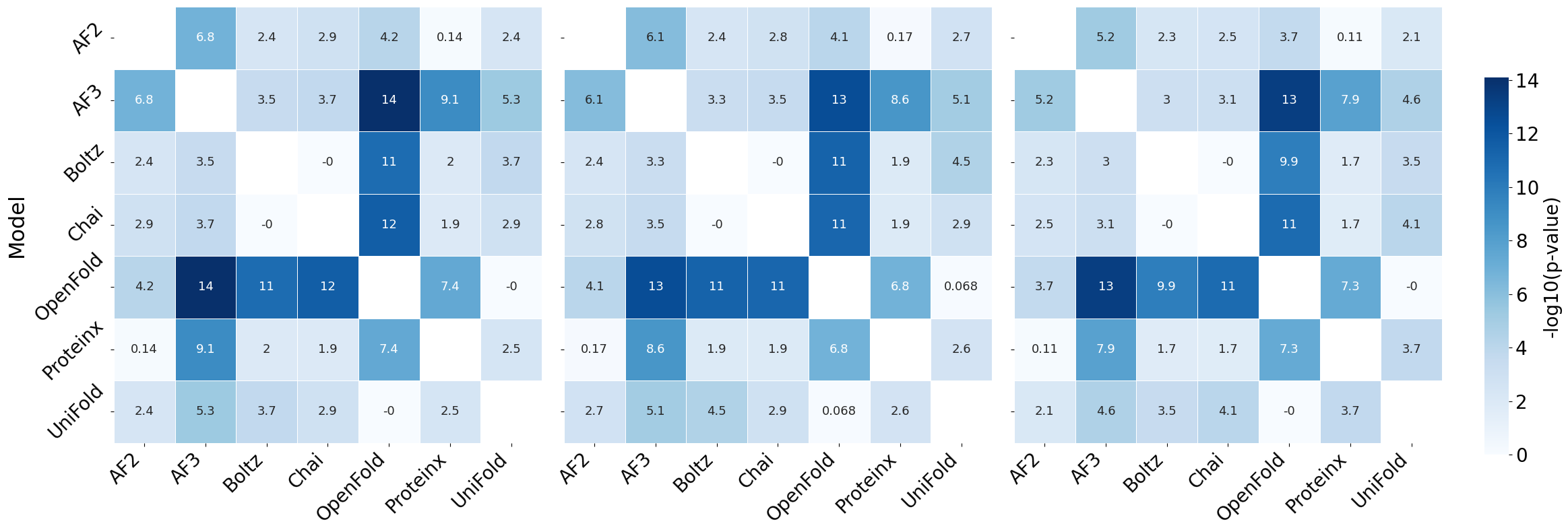}
    \caption{Heatmaps of $-\log_{10}(p)$ values from McNemar tests comparing pairwise model performance on PPI prediction across different $\mathrm{FUS}_T(\tau)$ thresholds. Left: full sequence; Middle: $\mathrm{FUS}_T(\tau=30)$; Right: $\mathrm{FUS}_T(\tau=50)$. Higher values indicate more statistical significance between PSPMs. Blank blocks indicate self-comparisons, which are omitted by definition.}
    \label{fig:pvalue_ppi}
    \vspace{-1em}
\end{figure*}
\section{Experimental Results and Analysis}
\label{sec:exp}
\begin{figure*}[h!]
    \centering   
    \includegraphics[width=\linewidth]{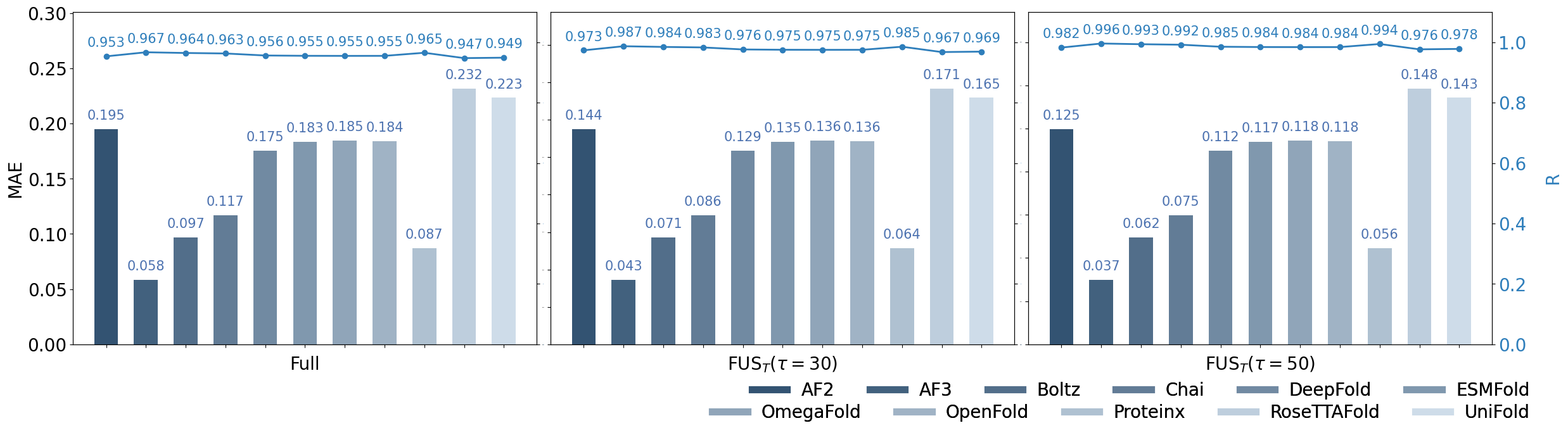}
    \caption{Performance comparison of MAE and $R$ on drug discovery prediction across PSPMs under different $\mathrm{FUS}_T(\tau)$ thresholds.}
    \label{fig:mae_r}
\end{figure*}
\begin{figure*}[h!]
    \centering
    \includegraphics[width=\linewidth]{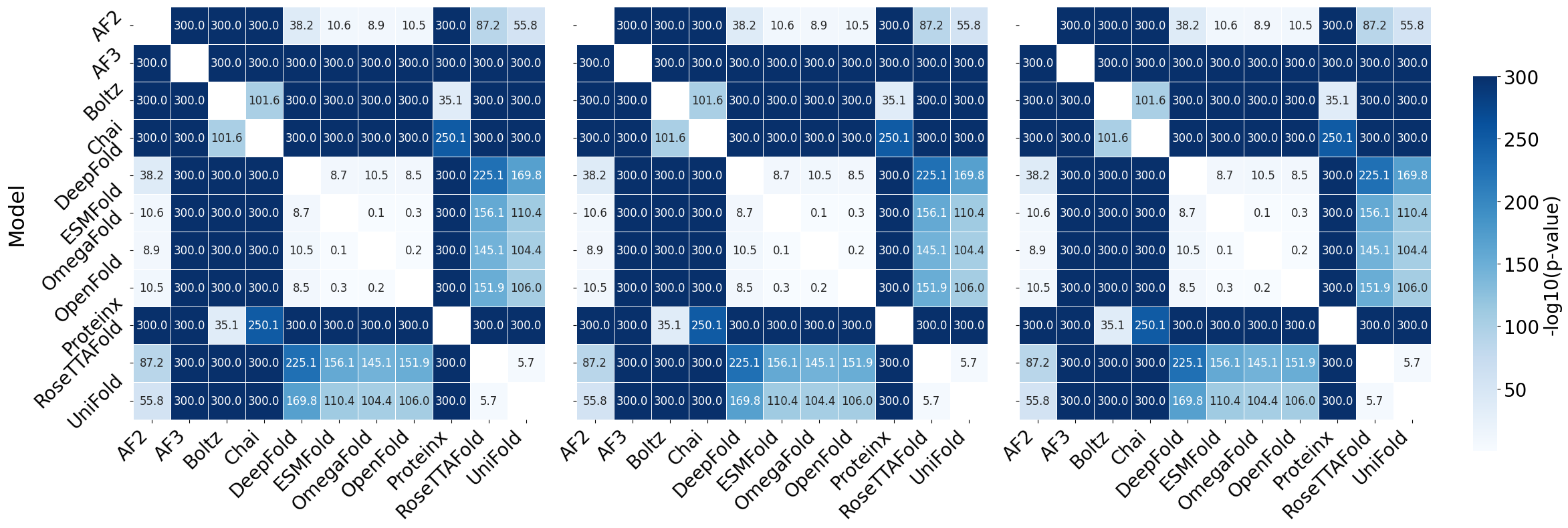}
    \caption{Heatmaps of $-\log_{10}(p)$ values from Wilcoxon signed-rank tests comparing model performance in drug discovery tasks across different $\mathrm{FUS}_T(\tau)$ thresholds. Left: full sequence; Middle: $\mathrm{FUS}_T(\tau=30)$; Right: $\mathrm{FUS}_T(\tau=50)$. Higher values indicate greater statistical significance in pairwise differences between PSPMs. Blank blocks indicate self-comparisons, which are omitted by definition.}
    \label{fig:pvalue_drug}
    \vspace{-1em}
\end{figure*}

After introducing the model suite and evaluation toolbox, we evaluate whether disorder-aware uncertainty manifests as task-dependent functional effects. Rather than ranking PSPMs by absolute accuracy, we ask whether uncertainty localized in intrinsically disordered regions (IDRs) propagates differently across downstream tasks.
We therefore study two tasks with contrasting sensitivity to disorder: \textit{(i) protein–protein interaction (PPI) prediction}, which depends on flexible, disorder-mediated interfaces, and \textbf{(ii) structure-based drug discovery}, where signal is concentrated in compact, well-structured binding pockets. If structural uncertainty were functionally interchangeable, uncertainty-stratified filtering would produce similar performance shifts in both tasks.

\subsection{Validation on PPI}
\label{sec:ppi}

We first evaluate the functional impact of structural uncertainty on PPI prediction. As shown in Table~\ref{tab:combined_ppi_metrics}, uncertainty-stratified  filtering using $\mathrm{FUS}_T(\tau)$ consistently improves accuracy, precision, recall, and F1 score across all evaluated PSPMs. These gains are modest but systematic, and increase with stricter filtering thresholds, indicating that IDRs introduce reproducible interface-level errors rather than random noise.

Importantly, the magnitude of improvement varies substantially across architectures. While models such as AF2, AF3, and ProteinX exhibit clear performance recovery under filtering, others (e.g., UniFold) show more limited gains despite comparable baseline performance. This heterogeneity suggests architecture-dependent robustness to IDR-driven uncertainty that is not captured by aggregate accuracy metrics alone.

Pairwise McNemar tests (Figure~\ref{fig:pvalue_ppi}) further confirm that stratified-stratified filtering induces statistically significant shifts in model behavior, and that these shifts differ across PSPMs. Together, these results indicate that structural uncertainty in IDRs has direct, task-relevant consequences for PPI prediction and that FUS reveals failure modes that are obscured in unfiltered evaluations.


\subsection{Validation on Drug Discovery}
\label{sec:drug}


In contrast, structure-based drug discovery exhibits limited sensitivity to uncertainty-stratified filtering. As shown in Figure~\ref{fig:mae_r}, both MAE and Pearson correlation remain largely unchanged across $\mathrm{FUS}_T(\tau)$ thresholds for all PSPMs. While absolute performance differs across architectures, within-model variation from the full set to $\tau=30$ and $\tau=50$ is minimal, with Pearson $R$ consistently near saturation and only minor, non-monotonic changes in MAE. This indicates that disorder-localized structural uncertainty has a weaker impact on this task compared to interaction-level predictions.

Wilcoxon signed-rank tests (Figure~\ref{fig:pvalue_drug}) corroborate this observation: while inter-model differences are often statistically significant, uncertainty-stratified filtering does not induce systematic within-model performance shifts. This pattern is consistent with the localization of task-relevant signal to compact binding pockets, which typically retain high structural uncertainty even in proteins containing extensive IDRs.

Together, these results validate the central premise of \model: structural uncertainty gives rise to task-dependent functional consequences rather than uniform degradation. PPI prediction is sensitive to IDR-driven uncertainty, with uncertainty-stratified filtering revealing architecture-dependent robustness and recovery effects. In contrast, structure-based drug discovery remains comparatively robust to the removal of IDRs. Crucially, this task-level divergence is largely invisible to aggregate accuracy metrics, underscoring the need for IDR-centric, uncertainty-aware diagnostics in benchmark design.

\section{Conclusion}
\label{sec:conclusion}

We introduced \model, an IDR-centric benchmark for evaluating the functional reliability of protein structure prediction models in biologically realistic settings. By explicitly incorporating structural uncertainty through uncertainty-stratified evaluation, the benchmark reveals task-dependent failure modes that are largely obscured by aggregate structural accuracy metrics. In particular, we show that IDRs substantially degrade PPI prediction, whereas structure-based drug discovery remains comparatively robust to structural uncertainty.

These findings expose fundamental limitations of one-size-fits-all evaluation practices and demonstrate the necessity of IDR-centric benchmarks for assessing structural AI systems. By releasing \model as an open, reproducible evaluation framework, we enable systematic diagnosis of model behavior under uncertainty and support principled comparison across models and tasks on IDRs. Future work includes extending \model to additional downstream tasks with different disorder profiles, incorporating dynamic or ensemble structural representations, and exploring alternative uncertainty estimates beyond model-reported uncertainty scores. These directions would further strengthen disorder-aware evaluation and broaden the applicability of FUS-based diagnostics.

\newpage

\bibliographystyle{ACM-Reference-Format} 
\bibliography{references}


\begin{thebibliography}{35}


\ifx \showCODEN    \undefined \def \showCODEN     #1{\unskip}     \fi
\ifx \showISBNx    \undefined \def \showISBNx     #1{\unskip}     \fi
\ifx \showISBNxiii \undefined \def \showISBNxiii  #1{\unskip}     \fi
\ifx \showISSN     \undefined \def \showISSN      #1{\unskip}     \fi
\ifx \showLCCN     \undefined \def \showLCCN      #1{\unskip}     \fi
\ifx \shownote     \undefined \def \shownote      #1{#1}          \fi
\ifx \showarticletitle \undefined \def \showarticletitle #1{#1}   \fi
\ifx \showURL      \undefined \def \showURL       {\relax}        \fi
\providecommand\bibfield[2]{#2}
\providecommand\bibinfo[2]{#2}
\providecommand\natexlab[1]{#1}
\providecommand\showeprint[2][]{arXiv:#2}

\bibitem[Abramson et~al\mbox{.}(2024)]%
        {abramson2024alphafold3}
\bibfield{author}{\bibinfo{person}{Jack Abramson}, \bibinfo{person}{Jonathan Adler}, \bibinfo{person}{James Dunger}, {et~al\mbox{.}}} \bibinfo{year}{2024}\natexlab{}.
\newblock \showarticletitle{Accurate structure prediction of biomolecular interactions with AlphaFold 3}.
\newblock \bibinfo{journal}{\emph{Nature}}  \bibinfo{volume}{630} (\bibinfo{year}{2024}), \bibinfo{pages}{493--500}.
\newblock
\href{https://doi.org/10.1038/s41586-024-07487-w}{doi:\nolinkurl{10.1038/s41586-024-07487-w}}


\bibitem[Ahdritz et~al\mbox{.}(2024)]%
        {Ahdritz2024OpenFold}
\bibfield{author}{\bibinfo{person}{Gustav Ahdritz}, \bibinfo{person}{Nazim Bouatta}, \bibinfo{person}{Cristian Floristean}, {et~al\mbox{.}}} \bibinfo{year}{2024}\natexlab{}.
\newblock \showarticletitle{OpenFold: retraining AlphaFold2 yields new insights into its learning mechanisms and capacity for generalization}.
\newblock \bibinfo{journal}{\emph{Nature Methods}}  \bibinfo{volume}{21} (\bibinfo{year}{2024}), \bibinfo{pages}{1514--1524}.
\newblock
\href{https://doi.org/10.1038/s41592-024-02272-z}{doi:\nolinkurl{10.1038/s41592-024-02272-z}}


\bibitem[Anfinsen(1973)]%
        {anfinsen1973principles}
\bibfield{author}{\bibinfo{person}{Christian~B Anfinsen}.} \bibinfo{year}{1973}\natexlab{}.
\newblock \showarticletitle{Principles that govern the folding of protein chains}.
\newblock \bibinfo{journal}{\emph{Science}} \bibinfo{volume}{181}, \bibinfo{number}{4096} (\bibinfo{year}{1973}), \bibinfo{pages}{223--230}.
\newblock


\bibitem[Baek et~al\mbox{.}(2021)]%
        {baek2021accurate}
\bibfield{author}{\bibinfo{person}{Minkyung Baek}, \bibinfo{person}{Frank DiMaio}, \bibinfo{person}{Ivan Anishchenko}, \bibinfo{person}{Justas Dauparas}, \bibinfo{person}{Sergey Ovchinnikov}, \bibinfo{person}{Gyu~Rie Lee}, \bibinfo{person}{Jue Wang}, \bibinfo{person}{Qian Cong}, \bibinfo{person}{Lisa~N Kinch}, \bibinfo{person}{R~Dustin Schaeffer}, {et~al\mbox{.}}} \bibinfo{year}{2021}\natexlab{}.
\newblock \showarticletitle{Accurate prediction of protein structures and interactions using a three-track neural network}.
\newblock \bibinfo{journal}{\emph{Science}} \bibinfo{volume}{373}, \bibinfo{number}{6557} (\bibinfo{year}{2021}), \bibinfo{pages}{871--876}.
\newblock


\bibitem[Bondos et~al\mbox{.}(2022)]%
        {bondos2022intrinsically}
\bibfield{author}{\bibinfo{person}{Sarah~E Bondos}, \bibinfo{person}{A~Keith Dunker}, {and} \bibinfo{person}{Vladimir~N Uversky}.} \bibinfo{year}{2022}\natexlab{}.
\newblock \showarticletitle{Intrinsically disordered proteins play diverse roles in cell signaling}.
\newblock \bibinfo{journal}{\emph{Cell Communication and Signaling}} \bibinfo{volume}{20}, \bibinfo{number}{1} (\bibinfo{year}{2022}), \bibinfo{pages}{20}.
\newblock


\bibitem[{Chai Discovery}(2024)]%
        {Chai-1-Technical-Report}
\bibfield{author}{\bibinfo{person}{{Chai Discovery}}.} \bibinfo{year}{2024}\natexlab{}.
\newblock \showarticletitle{Chai-1: Decoding the molecular interactions of life}.
\newblock \bibinfo{journal}{\emph{bioRxiv}} (\bibinfo{year}{2024}).
\newblock
\href{https://doi.org/10.1101/2024.10.10.615955}{doi:\nolinkurl{10.1101/2024.10.10.615955}}


\bibitem[Chang et~al\mbox{.}(2024)]%
        {chang2024revolutionizing}
\bibfield{author}{\bibinfo{person}{Liwei Chang}, \bibinfo{person}{Arup Mondal}, \bibinfo{person}{Bhumika Singh}, \bibinfo{person}{Yisel Mart{\'\i}nez-Noa}, {and} \bibinfo{person}{Alberto Perez}.} \bibinfo{year}{2024}\natexlab{}.
\newblock \showarticletitle{Revolutionizing peptide-based drug discovery: Advances in the post-AlphaFold era}.
\newblock \bibinfo{journal}{\emph{Wiley Interdisciplinary Reviews: Computational Molecular Science}} \bibinfo{volume}{14}, \bibinfo{number}{1} (\bibinfo{year}{2024}), \bibinfo{pages}{e1693}.
\newblock


\bibitem[Chen et~al\mbox{.}(2025)]%
        {chen2025protenix}
\bibfield{author}{\bibinfo{person}{Xinshi Chen}, \bibinfo{person}{Yuxuan Zhang}, \bibinfo{person}{Chan Lu}, \bibinfo{person}{Wenzhi Ma}, \bibinfo{person}{Jiaqi Guan}, \bibinfo{person}{Chengyue Gong}, \bibinfo{person}{Jincai Yang}, \bibinfo{person}{Hanyu Zhang}, \bibinfo{person}{Ke Zhang}, \bibinfo{person}{Shenghao Wu}, \bibinfo{person}{Kuangqi Zhou}, \bibinfo{person}{Yanping Yang}, \bibinfo{person}{Zhenyu Liu}, \bibinfo{person}{Lan Wang}, \bibinfo{person}{Bo Shi}, \bibinfo{person}{Shaochen Shi}, {and} \bibinfo{person}{Wenzhi Xiao}.} \bibinfo{year}{2025}\natexlab{}.
\newblock \showarticletitle{Protenix - Advancing Structure Prediction Through a Comprehensive AlphaFold3 Reproduction}.
\newblock \bibinfo{journal}{\emph{bioRxiv}} (\bibinfo{year}{2025}).
\newblock
\href{https://doi.org/10.1101/2025.01.08.631967}{doi:\nolinkurl{10.1101/2025.01.08.631967}}


\bibitem[Dhakal et~al\mbox{.}(2022)]%
        {dhakal2022artificial}
\bibfield{author}{\bibinfo{person}{Ashwin Dhakal}, \bibinfo{person}{Cole McKay}, \bibinfo{person}{John~J Tanner}, {and} \bibinfo{person}{Jianlin Cheng}.} \bibinfo{year}{2022}\natexlab{}.
\newblock \showarticletitle{Artificial intelligence in the prediction of protein--ligand interactions: recent advances and future directions}.
\newblock \bibinfo{journal}{\emph{Briefings in bioinformatics}} \bibinfo{volume}{23}, \bibinfo{number}{1} (\bibinfo{year}{2022}), \bibinfo{pages}{bbab476}.
\newblock


\bibitem[Dill et~al\mbox{.}(2008)]%
        {dill2008protein}
\bibfield{author}{\bibinfo{person}{Ken~A Dill}, \bibinfo{person}{S~Banu Ozkan}, \bibinfo{person}{M~Scott Shell}, {and} \bibinfo{person}{Thomas~R Weikl}.} \bibinfo{year}{2008}\natexlab{}.
\newblock \showarticletitle{The protein folding problem}.
\newblock \bibinfo{journal}{\emph{Annu. Rev. Biophys.}} \bibinfo{volume}{37}, \bibinfo{number}{1} (\bibinfo{year}{2008}), \bibinfo{pages}{289--316}.
\newblock


\bibitem[Erd{\H{o}}s and Doszt{\'a}nyi(2024)]%
        {erdHos2024deep}
\bibfield{author}{\bibinfo{person}{G{\'a}bor Erd{\H{o}}s} {and} \bibinfo{person}{Zsuzsanna Doszt{\'a}nyi}.} \bibinfo{year}{2024}\natexlab{}.
\newblock \showarticletitle{Deep learning for intrinsically disordered proteins: From improved predictions to deciphering conformational ensembles}.
\newblock \bibinfo{journal}{\emph{Current Opinion in Structural Biology}}  \bibinfo{volume}{89} (\bibinfo{year}{2024}), \bibinfo{pages}{102950}.
\newblock


\bibitem[Evans et~al\mbox{.}(2021)]%
        {evans2021protein}
\bibfield{author}{\bibinfo{person}{Richard Evans}, \bibinfo{person}{Michael O’Neill}, \bibinfo{person}{Alexander Pritzel}, \bibinfo{person}{Natasha Antropova}, \bibinfo{person}{Andrew Senior}, \bibinfo{person}{Tim Green}, \bibinfo{person}{Augustin {\v{Z}}{\'\i}dek}, \bibinfo{person}{Russ Bates}, \bibinfo{person}{Sam Blackwell}, \bibinfo{person}{Jason Yim}, {et~al\mbox{.}}} \bibinfo{year}{2021}\natexlab{}.
\newblock \showarticletitle{Protein complex prediction with AlphaFold-Multimer}.
\newblock \bibinfo{journal}{\emph{biorxiv}} (\bibinfo{year}{2021}), \bibinfo{pages}{2021--10}.
\newblock


\bibitem[Hu and Ohue(2024)]%
        {hu2024spatialppi}
\bibfield{author}{\bibinfo{person}{Wenxing Hu} {and} \bibinfo{person}{Masahito Ohue}.} \bibinfo{year}{2024}\natexlab{}.
\newblock \showarticletitle{SpatialPPI: Three-dimensional space protein-protein interaction prediction with AlphaFold Multimer}.
\newblock \bibinfo{journal}{\emph{Computational and Structural Biotechnology Journal}}  \bibinfo{volume}{23} (\bibinfo{year}{2024}), \bibinfo{pages}{1214--1225}.
\newblock


\bibitem[Jumper et~al\mbox{.}(2021)]%
        {jumper2021highly}
\bibfield{author}{\bibinfo{person}{John Jumper}, \bibinfo{person}{Richard Evans}, \bibinfo{person}{Alexander Pritzel}, \bibinfo{person}{Tim Green}, \bibinfo{person}{Michael Figurnov}, \bibinfo{person}{Olaf Ronneberger}, \bibinfo{person}{Kathryn Tunyasuvunakool}, \bibinfo{person}{Russ Bates}, \bibinfo{person}{Augustin {\v{Z}}{\'\i}dek}, \bibinfo{person}{Anna Potapenko}, {et~al\mbox{.}}} \bibinfo{year}{2021}\natexlab{}.
\newblock \showarticletitle{Highly accurate protein structure prediction with AlphaFold}.
\newblock \bibinfo{journal}{\emph{nature}} \bibinfo{volume}{596}, \bibinfo{number}{7873} (\bibinfo{year}{2021}), \bibinfo{pages}{583--589}.
\newblock


\bibitem[Krokidis et~al\mbox{.}(2025)]%
        {krokidis2025alphafold3}
\bibfield{author}{\bibinfo{person}{Marios~G Krokidis}, \bibinfo{person}{Dimitrios~E Koumadorakis}, \bibinfo{person}{Konstantinos Lazaros}, \bibinfo{person}{Ouliana Ivantsik}, \bibinfo{person}{Themis~P Exarchos}, \bibinfo{person}{Aristidis~G Vrahatis}, \bibinfo{person}{Sotiris Kotsiantis}, {and} \bibinfo{person}{Panagiotis Vlamos}.} \bibinfo{year}{2025}\natexlab{}.
\newblock \showarticletitle{AlphaFold3: an overview of applications and performance insights}.
\newblock \bibinfo{journal}{\emph{International Journal of Molecular Sciences}} \bibinfo{volume}{26}, \bibinfo{number}{8} (\bibinfo{year}{2025}), \bibinfo{pages}{3671}.
\newblock


\bibitem[Lee et~al\mbox{.}(2023)]%
        {Lee2023DeepFold}
\bibfield{author}{\bibinfo{person}{J.~W. Lee}, \bibinfo{person}{J.~H. Won}, \bibinfo{person}{S. Jeon}, \bibinfo{person}{Y. Choo}, \bibinfo{person}{Y. Yeon}, \bibinfo{person}{J.~S. Oh}, \bibinfo{person}{M. Kim}, \bibinfo{person}{S. Kim}, \bibinfo{person}{I. Joung}, \bibinfo{person}{C. Jang}, \bibinfo{person}{S.~J. Lee}, \bibinfo{person}{T.~H. Kim}, \bibinfo{person}{K.~H. Jin}, \bibinfo{person}{G. Song}, \bibinfo{person}{E.~S. Kim}, \bibinfo{person}{J. Yoo}, \bibinfo{person}{E. Paek}, \bibinfo{person}{Y.~K. Noh}, {and} \bibinfo{person}{K. Joo}.} \bibinfo{year}{2023}\natexlab{}.
\newblock \showarticletitle{DeepFold: enhancing protein structure prediction through optimized loss functions, improved template features, and re-optimized energy function}.
\newblock \bibinfo{journal}{\emph{Bioinformatics}} \bibinfo{volume}{39}, \bibinfo{number}{12} (\bibinfo{year}{2023}), \bibinfo{pages}{btad712}.
\newblock
\href{https://doi.org/10.1093/bioinformatics/btad712}{doi:\nolinkurl{10.1093/bioinformatics/btad712}}


\bibitem[Li et~al\mbox{.}(2022)]%
        {Li2022UniFold}
\bibfield{author}{\bibinfo{person}{Ziyao Li}, \bibinfo{person}{Xuyang Liu}, \bibinfo{person}{Weijie Chen}, \bibinfo{person}{Fan Shen}, \bibinfo{person}{Hangrui Bi}, \bibinfo{person}{Guolin Ke}, {and} \bibinfo{person}{Linfeng Zhang}.} \bibinfo{year}{2022}\natexlab{}.
\newblock \showarticletitle{Uni-Fold: An Open-Source Platform for Developing Protein Folding Models beyond AlphaFold}.
\newblock \bibinfo{journal}{\emph{bioRxiv}} (\bibinfo{year}{2022}).
\newblock
\href{https://doi.org/10.1101/2022.08.04.502811}{doi:\nolinkurl{10.1101/2022.08.04.502811}}
\newblock
\shownote{preprint}.


\bibitem[Lin et~al\mbox{.}(2023)]%
        {lin2023evolutionary}
\bibfield{author}{\bibinfo{person}{Zeming Lin}, \bibinfo{person}{Halil Akin}, \bibinfo{person}{Roshan Rao}, \bibinfo{person}{Brian Hie}, \bibinfo{person}{Zhongkai Zhu}, \bibinfo{person}{Wenting Lu}, \bibinfo{person}{Nikita Smetanin}, \bibinfo{person}{Robert Verkuil}, \bibinfo{person}{Ori Kabeli}, \bibinfo{person}{Yaniv Shmueli}, {et~al\mbox{.}}} \bibinfo{year}{2023}\natexlab{}.
\newblock \showarticletitle{Evolutionary-scale prediction of atomic-level protein structure with a language model}.
\newblock \bibinfo{journal}{\emph{Science}} \bibinfo{volume}{379}, \bibinfo{number}{6637} (\bibinfo{year}{2023}), \bibinfo{pages}{1123--1130}.
\newblock


\bibitem[Lotthammer et~al\mbox{.}(2024)]%
        {lotthammer2024direct}
\bibfield{author}{\bibinfo{person}{Jeffrey~M Lotthammer}, \bibinfo{person}{Garrett~M Ginell}, \bibinfo{person}{Daniel Griffith}, \bibinfo{person}{Ryan Emenecker}, {and} \bibinfo{person}{Alex~S Holehouse}.} \bibinfo{year}{2024}\natexlab{}.
\newblock \showarticletitle{Direct prediction of intrinsically disordered protein conformational properties from sequence}.
\newblock \bibinfo{journal}{\emph{Biophysical Journal}} \bibinfo{volume}{123}, \bibinfo{number}{3} (\bibinfo{year}{2024}), \bibinfo{pages}{43a}.
\newblock


\bibitem[Madhurima et~al\mbox{.}(2023)]%
        {madhurima2023functional}
\bibfield{author}{\bibinfo{person}{Kulkarni Madhurima}, \bibinfo{person}{Bodhisatwa Nandi}, \bibinfo{person}{Sneha Munshi}, \bibinfo{person}{Athi~N Naganathan}, {and} \bibinfo{person}{Ashok Sekhar}.} \bibinfo{year}{2023}\natexlab{}.
\newblock \showarticletitle{Functional regulation of an intrinsically disordered protein via a conformationally excited state}.
\newblock \bibinfo{journal}{\emph{Science advances}} \bibinfo{volume}{9}, \bibinfo{number}{26} (\bibinfo{year}{2023}), \bibinfo{pages}{eadh4591}.
\newblock


\bibitem[Moult et~al\mbox{.}(2018)]%
        {moult2018critical}
\bibfield{author}{\bibinfo{person}{John Moult}, \bibinfo{person}{Krzysztof Fidelis}, \bibinfo{person}{Andriy Kryshtafovych}, \bibinfo{person}{Torsten Schwede}, {and} \bibinfo{person}{Anna Tramontano}.} \bibinfo{year}{2018}\natexlab{}.
\newblock \showarticletitle{Critical assessment of methods of protein structure prediction (CASP)—Round XII}.
\newblock \bibinfo{journal}{\emph{Proteins: Structure, Function, and Bioinformatics}}  \bibinfo{volume}{86} (\bibinfo{year}{2018}), \bibinfo{pages}{7--15}.
\newblock


\bibitem[Moult et~al\mbox{.}(1995)]%
        {moult1995large}
\bibfield{author}{\bibinfo{person}{John Moult}, \bibinfo{person}{Jan~T Pedersen}, \bibinfo{person}{Richard Judson}, {and} \bibinfo{person}{Krzysztof Fidelis}.} \bibinfo{year}{1995}\natexlab{}.
\newblock \bibinfo{title}{A large-scale experiment to assess protein structure prediction methods}.
\newblock \bibinfo{numpages}{ii--iv}~pages.
\newblock


\bibitem[Oldfield and Dunker(2014)]%
        {oldfield2014intrinsically}
\bibfield{author}{\bibinfo{person}{Christopher~J Oldfield} {and} \bibinfo{person}{A~Keith Dunker}.} \bibinfo{year}{2014}\natexlab{}.
\newblock \showarticletitle{Intrinsically disordered proteins and intrinsically disordered protein regions}.
\newblock \bibinfo{journal}{\emph{Annual review of biochemistry}} \bibinfo{volume}{83}, \bibinfo{number}{1} (\bibinfo{year}{2014}), \bibinfo{pages}{553--584}.
\newblock


\bibitem[Piovesan et~al\mbox{.}(2022)]%
        {piovesan2022intrinsic}
\bibfield{author}{\bibinfo{person}{Damiano Piovesan}, \bibinfo{person}{Alexander~Miguel Monzon}, {and} \bibinfo{person}{Silvio~CE Tosatto}.} \bibinfo{year}{2022}\natexlab{}.
\newblock \showarticletitle{Intrinsic protein disorder and conditional folding in AlphaFoldDB}.
\newblock \bibinfo{journal}{\emph{Protein Science}} \bibinfo{volume}{31}, \bibinfo{number}{11} (\bibinfo{year}{2022}), \bibinfo{pages}{e4466}.
\newblock


\bibitem[Rohl et~al\mbox{.}(2004)]%
        {rohl2004protein}
\bibfield{author}{\bibinfo{person}{Carol~A Rohl}, \bibinfo{person}{Charlie~EM Strauss}, \bibinfo{person}{Kira~MS Misura}, {and} \bibinfo{person}{David Baker}.} \bibinfo{year}{2004}\natexlab{}.
\newblock \showarticletitle{Protein structure prediction using Rosetta}.
\newblock In \bibinfo{booktitle}{\emph{Methods in enzymology}}. Vol.~\bibinfo{volume}{383}. \bibinfo{publisher}{Elsevier}, \bibinfo{pages}{66--93}.
\newblock


\bibitem[Trivedi and Nagarajaram(2022)]%
        {trivedi2022intrinsically}
\bibfield{author}{\bibinfo{person}{Rakesh Trivedi} {and} \bibinfo{person}{Hampapathalu~Adimurthy Nagarajaram}.} \bibinfo{year}{2022}\natexlab{}.
\newblock \showarticletitle{Intrinsically disordered proteins: an overview}.
\newblock \bibinfo{journal}{\emph{International journal of molecular sciences}} \bibinfo{volume}{23}, \bibinfo{number}{22} (\bibinfo{year}{2022}), \bibinfo{pages}{14050}.
\newblock


\bibitem[Uversky(2019)]%
        {uversky2019intrinsically}
\bibfield{author}{\bibinfo{person}{Vladimir~N Uversky}.} \bibinfo{year}{2019}\natexlab{}.
\newblock \showarticletitle{Intrinsically disordered proteins and their “mysterious”(meta) physics}.
\newblock \bibinfo{journal}{\emph{Frontiers in Physics}}  \bibinfo{volume}{7} (\bibinfo{year}{2019}), \bibinfo{pages}{10}.
\newblock


\bibitem[Vangone and Bonvin(2015)]%
        {vangone2015contacts}
\bibfield{author}{\bibinfo{person}{Anna Vangone} {and} \bibinfo{person}{Alexandre~MJJ Bonvin}.} \bibinfo{year}{2015}\natexlab{}.
\newblock \showarticletitle{Contacts-based prediction of binding affinity in protein--protein complexes}.
\newblock \bibinfo{journal}{\emph{elife}}  \bibinfo{volume}{4} (\bibinfo{year}{2015}), \bibinfo{pages}{e07454}.
\newblock


\bibitem[Verburgt et~al\mbox{.}(2022)]%
        {Verburgt2022}
\bibfield{author}{\bibinfo{person}{J. Verburgt}, \bibinfo{person}{Z. Zhang}, {and} \bibinfo{person}{D. Kihara}.} \bibinfo{year}{2022}\natexlab{}.
\newblock \showarticletitle{Multi-level analysis of intrinsically disordered protein docking methods}.
\newblock \bibinfo{journal}{\emph{Methods}}  \bibinfo{volume}{204} (\bibinfo{year}{2022}), \bibinfo{pages}{55--63}.
\newblock
\href{https://doi.org/10.1016/j.ymeth.2022.05.006}{doi:\nolinkurl{10.1016/j.ymeth.2022.05.006}}
\newblock
\shownote{San Diego, Calif.}.


\bibitem[Wohlwend et~al\mbox{.}(2024)]%
        {wohlwend2024boltz1}
\bibfield{author}{\bibinfo{person}{Jeremy Wohlwend}, \bibinfo{person}{Gabriele Corso}, \bibinfo{person}{Saro Passaro}, \bibinfo{person}{Noah Getz}, \bibinfo{person}{Mateo Reveiz}, \bibinfo{person}{Ken Leidal}, \bibinfo{person}{Wojtek Swiderski}, \bibinfo{person}{Liam Atkinson}, \bibinfo{person}{Tally Portnoi}, \bibinfo{person}{Itamar Chinn}, \bibinfo{person}{Jacob Silterra}, \bibinfo{person}{Tommi Jaakkola}, {and} \bibinfo{person}{Regina Barzilay}.} \bibinfo{year}{2024}\natexlab{}.
\newblock \showarticletitle{Boltz-1: Democratizing Biomolecular Interaction Modeling}.
\newblock \bibinfo{journal}{\emph{bioRxiv}} (\bibinfo{year}{2024}).
\newblock
\href{https://doi.org/10.1101/2024.11.19.624167}{doi:\nolinkurl{10.1101/2024.11.19.624167}}


\bibitem[Wright and Dyson(2015)]%
        {wright2015intrinsically}
\bibfield{author}{\bibinfo{person}{Peter~E Wright} {and} \bibinfo{person}{H~Jane Dyson}.} \bibinfo{year}{2015}\natexlab{}.
\newblock \showarticletitle{Intrinsically disordered proteins in cellular signalling and regulation}.
\newblock \bibinfo{journal}{\emph{Nature reviews Molecular cell biology}} \bibinfo{volume}{16}, \bibinfo{number}{1} (\bibinfo{year}{2015}), \bibinfo{pages}{18--29}.
\newblock


\bibitem[Wu et~al\mbox{.}(2022)]%
        {OmegaFold}
\bibfield{author}{\bibinfo{person}{Ruidong Wu}, \bibinfo{person}{Fan Ding}, \bibinfo{person}{Rui Wang}, \bibinfo{person}{Rui Shen}, \bibinfo{person}{Xiwen Zhang}, \bibinfo{person}{Shitong Luo}, \bibinfo{person}{Chenpeng Su}, \bibinfo{person}{Zuofan Wu}, \bibinfo{person}{Qi Xie}, \bibinfo{person}{Bonnie Berger}, \bibinfo{person}{Jianzhu Ma}, {and} \bibinfo{person}{Jian Peng}.} \bibinfo{year}{2022}\natexlab{}.
\newblock \showarticletitle{High-resolution de novo structure prediction from primary sequence}.
\newblock \bibinfo{journal}{\emph{bioRxiv}} (\bibinfo{year}{2022}).
\newblock
\href{https://doi.org/10.1101/2022.07.21.500999}{doi:\nolinkurl{10.1101/2022.07.21.500999}}


\bibitem[Yang et~al\mbox{.}(2024)]%
        {yang2024deep}
\bibfield{author}{\bibinfo{person}{Yuxin Yang}, \bibinfo{person}{Yunguang Qiu}, \bibinfo{person}{Jianying Hu}, \bibinfo{person}{Michal Rosen-Zvi}, \bibinfo{person}{Qiang Guan}, {and} \bibinfo{person}{Feixiong Cheng}.} \bibinfo{year}{2024}\natexlab{}.
\newblock \showarticletitle{A deep learning framework combining molecular image and protein structural representations identifies candidate drugs for pain}.
\newblock \bibinfo{journal}{\emph{Cell Reports Methods}} \bibinfo{volume}{4}, \bibinfo{number}{10} (\bibinfo{year}{2024}).
\newblock


\bibitem[Zhang et~al\mbox{.}(2024)]%
        {zhang2024machine}
\bibfield{author}{\bibinfo{person}{Yunjiang Zhang}, \bibinfo{person}{Shuyuan Li}, \bibinfo{person}{Kong Meng}, {and} \bibinfo{person}{Shaorui Sun}.} \bibinfo{year}{2024}\natexlab{}.
\newblock \showarticletitle{Machine Learning for Sequence and Structure-Based Protein--Ligand Interaction Prediction}.
\newblock \bibinfo{journal}{\emph{Journal of chemical information and modeling}} \bibinfo{volume}{64}, \bibinfo{number}{5} (\bibinfo{year}{2024}), \bibinfo{pages}{1456--1472}.
\newblock


\bibitem[Zhang et~al\mbox{.}(2025)]%
        {zhang2025protein}
\bibfield{author}{\bibinfo{person}{Yumeng Zhang}, \bibinfo{person}{Jared Zheng}, {and} \bibinfo{person}{Bin Zhang}.} \bibinfo{year}{2025}\natexlab{}.
\newblock \showarticletitle{Protein language model identifies disordered, conserved motifs implicated in phase separation}.
\newblock \bibinfo{journal}{\emph{Elife}}  \bibinfo{volume}{14} (\bibinfo{year}{2025}), \bibinfo{pages}{RP105309}.
\newblock


\end{thebibliography}

\newpage
\section*{Appendix}
\label{append:exp}
\subsection*{1. Evaluation Setup}

To account for structural uncertainty in a task-aware manner, we stratify all evaluations using \textit{Functional Uncertainty Sensitivity(FUS)}, denoted as $\mathrm{FUS}_T(\tau)$. Rather than relying solely on raw per-residue confidence scores, $\mathrm{FUS}_T(\tau)$ filters structural regions based on their estimated contribution to downstream functional tasks, explicitly capturing the interaction between disorder, confidence, and task relevance.

We report results under three regimes: the full sequence (no filtering), moderately constrained regions $\mathrm{FUS}_T(\tau{=}30)$, and highly constrained regions $\mathrm{FUS}_T(\tau{=}50)$. This stratification enables systematic analysis of model robustness as progressively more ambiguous and disorder-dominated regions are excluded, while preserving biologically meaningful interface information.

\subsection*{2. Overview of Benchmarked Model Architectures}

We benchmark 11 PSPMs and list the details of each model used in our benchmark. Each model is evaluated under uniform settings for protein–protein interaction (PPI) and/or drug discovery prediction tasks. This benchmarking effort specifically investigates how different model architectures perform in the presence of Intrinsically Disordered Regions (IDRs), which pose unique challenges for structural prediction and are critical for understanding flexible, functionally important protein interfaces.

\begin{itemize}
    \item \textbf{AlphaFold2 (AF2)}\cite{jumper2021highly}, developed by DeepMind, is built on the Evoformer architecture and structure module. The Evoformer processes both MSA and pairwise residue representations using a combination of axial attention, outer product mean, and triangle multiplication updates, enabling the modeling of long-range evolutionary dependencies. The structure module then predicts 3D atomic coordinates via invariant point attention and torsion angle regression. This architecture has set the standard for atomic-level protein structure prediction.
    \item \textbf{OpenFold}\cite{Ahdritz2024OpenFold}, developed by AQLab, is a faithful reimplementation of AF2 in PyTorch, retaining the Evoformer backbone and model design. While architecturally aligned with AF2, OpenFold emphasizes modularity and accessibility for the research community, enabling flexible adaptation and experimentation.
    \item \textbf{UniFold}\cite{Li2022UniFold}, introduced by DeepModeling, also replicates the AF2 architecture but incorporates computational optimizations for scalability. It supports distributed training and batched inference, allowing efficient deployment on large-scale protein datasets while maintaining the original Evoformer-based modeling capabilities.
    \item \textbf{AlphaFold3 (AF3)}\cite{abramson2024alphafold3}, the next-generation model from DeepMind, extends the Evoformer with a pretrained protein language model (LLM), enabling the integration of both MSA-derived and single-sequence embeddings. This multimodal approach enhances the model’s ability to generalize to MSA-scarce proteins. Additionally, AF3 introduces modules for ligand-aware modeling, supporting the prediction of protein-ligand complexes with atomic resolution. The model accepts both sequence and MSA inputs and outputs include 3D structures of proteins and bound ligands.
    \item \textbf{Boltz}\cite{wohlwend2024boltz1}, developed by the MIT Jameel Lab, is a transformer-based model that operates solely on sequence inputs. Its architecture consists of standard multi-head self-attention layers, trained with masked language modeling and supervised by structural signals such as contact maps. It is particularly suited for modeling flexible and disordered protein regions and produces coarse-grained structural outputs.
    \item \textbf{Chai}\cite{Chai-1-Technical-Report}, from Chai Discovery, uses a lightweight transformer architecture optimized for drug discovery. It focuses on capturing key functional sites with efficient attention mechanisms and positional encodings tailored to domain boundaries. The model operates on single-sequence inputs and is designed for rapid evaluation in therapeutic settings.
    \item \textbf{Proteinix}\cite{chen2025protenix}, introduced by ByteDance, enhances the transformer backbone with ligand-aware capabilities. It integrates structural cues from ligands during training and performs joint modeling of protein structures and ligand binding. The model outputs both atomic protein structures and ligand poses, enabling end-to-end prediction of molecular complexes.
    \item \textbf{ESMFold}\cite{lin2023evolutionary}, released by Meta AI, combines large-scale protein language modeling with a folding head to generate 3D structures. It uses the ESM-2 model to produce contextualized residue embeddings from single sequences, bypassing the need for MSA. These embeddings are then passed to a structure module that outputs backbone and side-chain coordinates. ESMFold offers exceptional speed and is especially useful for predicting structures of proteins without homologs.
    \item \textbf{OmegaFold}\cite{OmegaFold}, developed by HeliXon, uses a transformer-based encoder to embed sequence inputs and predict 3D coordinates through geometric vector outputs and torsion angle heads. It is designed for fast, MSA-free inference and supports coarse-grained output suitable for high-throughput applications in structural biology and drug discovery.
    \item \textbf{RoseTTAFold}\cite{baek2021accurate}, developed by the Baker Lab, employs a hybrid 3-track network that simultaneously processes 1D sequence features, 2D residue-residue relationships, and 3D coordinates. It combines convolutional layers for capturing local sequence context with attention mechanisms that model long-range interactions. Cross-track communication enables holistic structure learning, and the model predicts full atomic structures using MSA inputs. RoseTTAFold offers a fast alternative to AF2 with broad utility across structural tasks.
    \item \textbf{DeepFold}\cite{Lee2023DeepFold}, from Hanyang University, adopts a custom deep learning architecture comprising 1D convolutional layers, fully connected modules, and a geometric decoder. It operates solely on single-sequence inputs and is optimized for atomic-level predictions in drug-related tasks. The design emphasizes speed and simplicity while maintaining sufficient accuracy for practical applications in structure-based screening.
\end{itemize}

All models are evaluated under identical $\mathrm{FUS}_T(\tau)$ regimes to ensure fair comparison across architectures with different inductive biases toward disorder, flexibility, and interface modeling.

\subsection*{3. Downstream Applications and Evaluation Tasks}

Motivated by the limitations discussed in Section~\ref{sec:prelim}, we evaluate each PSPM on two key downstream tasks: PPI prediction and drug binding affinity estimation. These tasks require accurate structural modeling, particularly in regions identified as high-uncertainty under $\mathrm{FUS}_T(\tau)$, which frequently coincide with intrinsically disordered or conformationally flexible interface regions. This allows us to assess real-world utility beyond static structure accuracy.

\begin{figure}
    \centering
    \includegraphics[width=1\linewidth]{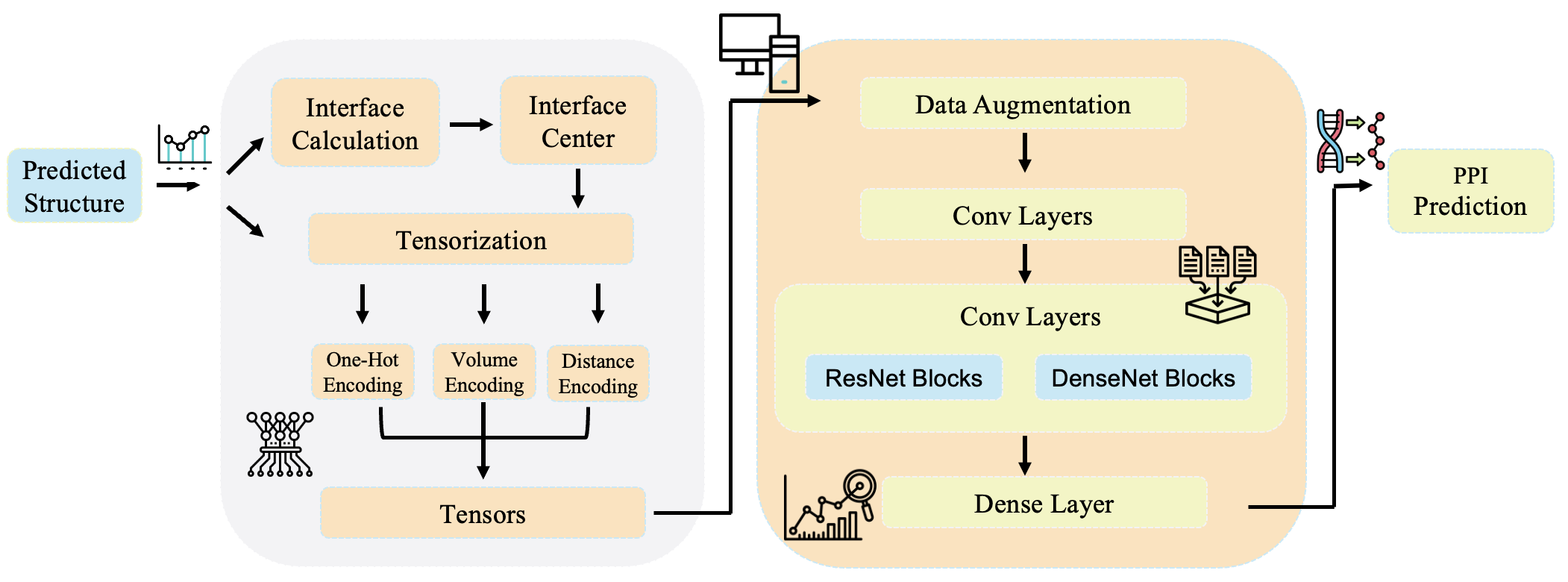}
    \caption{Overview of PPI Prediction}
    \label{fig:overview_ppi}
\end{figure}

\textbf{1. PPI Prediction} The pipeline begins with a predicted protein complex structure generated by the corresponding PSPM (e.g., AlphaFold-Multimer for AF2/AF3-based models). The resulting PDB file is used as input to extract the protein–protein interface, which is then encoded using one-hot, volumetric, or distance-based representations. This encoded tensor is passed through a neural network, where data augmentation is applied. Finally, a backbone architecture—either DenseNet or ResNet—predicts the probability that the input proteins interact(as shown in Figure~\ref{fig:overview_ppi}).

\begin{figure}
    \centering
    \includegraphics[width=1\linewidth]{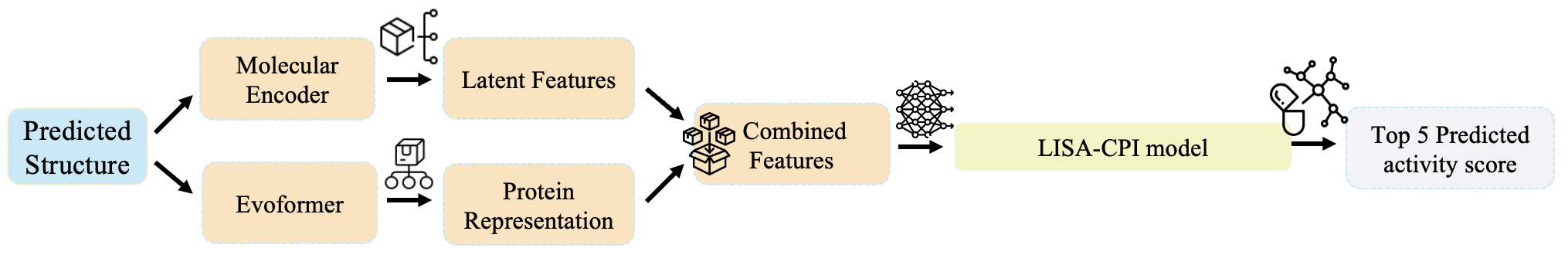}
    \caption{Overview of Drug Discovery }
    \label{fig:overview_drug}
\end{figure}

\textbf{2. Drug Discovery}
The pipeline begins with a predicted protein–ligand complex structure, which serves as input to two parallel modules. First, the molecular encoder processes the ligand component to extract latent chemical features. Simultaneously, the protein component is passed through an Evoformer module to derive a structured protein representation. These two feature sets are then concatenated to form a unified representation of the protein–ligand interaction. This combined feature vector is subsequently fed into the LISA-CPI model, which outputs a ranked list of predicted activity scores, with the top 5 scores indicating the most likely compound–protein interactions(as shown in Figure~\ref{fig:overview_drug}).

\subsection*{4. \model Experimental Design and Evaluation Framework}
To ensure robust and comprehensive evaluation, we conduct an extensive set of experiments within \model, a benchmark explicitly designed around functional uncertainty rather than static structural confidence. By systematically varying $\mathrm{FUS}_T(\tau)$ thresholds, evaluation metrics, and downstream task settings, we expose performance differences that are obscured when models are evaluated solely on high-confidence or well-folded regions.
This design enables us to characterize model sensitivity to uncertainty, disorder, and interface ambiguity in a unified framework, yielding a more realistic assessment of PSPM behavior in biologically relevant settings.

\subsubsection*{4.1 Definitions and Roles of Ligand and Receptor}

We follow the definitions of ligand and receptor in \cite{Verburgt2022}. The following definitions are used to compute structural interface scores, enabling consistent and biologically grounded evaluation across our benchmark.

\begin{figure}[h]
    \centering
    \includegraphics[width=1\linewidth]{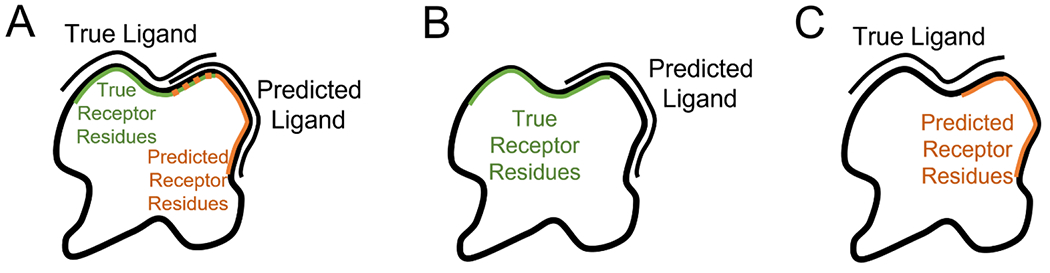}
    \caption{Diagrams depicting the interface residue metrics. A, Visualization of interfaces used in the Receptor Precision and Receptor Recall metrics. B, Visualization of the ligand and receptor residues used in the Ligand Precision metric. C, Visualization of the ligand and receptor residues used in the Ligand Recall metric.}
    \label{fig:ligand-receptor}
\end{figure}

\subsubsection*{4.2 Evaluation Metrics (Classification, Regression, and Structural)}
To comprehensively assess model performance across diverse interaction scenarios, we employ a suite of evaluation metrics that capture both predictive accuracy and structural fidelity. These metrics span classification, regression, and structure-aware categories, enabling robust comparison across tasks such as protein–protein interaction prediction and ligand binding affinity estimation. Below, we define each metric used in our benchmark and outline its relevance to the specific evaluation context.

\begin{table}[h!]
\centering
\renewcommand{\arraystretch}{2}
\begin{tabular}{p{4.5cm} p{4cm}}
\toprule
\textbf{Metric} & \textbf{Definition / Formula} \\
\midrule
Precision (Positive Predictive Value) &
$\displaystyle \frac{\text{TP}}{\text{TP} + \text{FP}}$ \\

Recall (Sensitivity) &
$\displaystyle \frac{\text{TP}}{\text{TP} + \text{FN}}$ \\

F1 Score &
$\displaystyle \frac{2 \cdot \text{TP}}{2 \cdot \text{TP} + \text{FP} + \text{FN}}$ \\

Accuracy &
$\displaystyle \frac{\text{TP} + \text{TN}}{\text{TP} + \text{TN} + \text{FP} + \text{FN}}$ \\
\bottomrule
\end{tabular}
\caption{Classification Metrics}
\end{table}

\begin{table}[h!]
\centering
\renewcommand{\arraystretch}{2}
\begin{tabular}{p{3.5cm} p{5cm}}
\toprule
\textbf{Metric} & \textbf{Definition / Formula} \\
\midrule
MAE &
$\displaystyle \frac{1}{N} \sum_{i=1}^{N} \left| y_i - \hat{y}_i \right|$ \\

MSE &
$\displaystyle \frac{1}{N} \sum_{i=1}^{N} \left( y_i - \hat{y}_i \right)^2$ \\

Pearson $R$ &
$\displaystyle \frac{ \sum_{i=1}^{N} (y_i - \bar{y})(\hat{y}_i - \bar{\hat{y}}) }
{ \sqrt{ \sum_{i=1}^{N} (y_i - \bar{y})^2 } \cdot \sqrt{ \sum_{i=1}^{N} (\hat{y}_i - \bar{\hat{y}})^2 } }$ \\
\bottomrule
\end{tabular}
\caption{Regression Metrics}
\end{table}

\begin{table}[h!]
\centering
\renewcommand{\arraystretch}{2}
\begin{tabular}{p{3cm} p{5.5cm}}
\toprule
\textbf{Metric} & \textbf{Definition / Formula} \\
\midrule
Receptor Precision (RP) &
{\tiny $\displaystyle \frac{|\text{True Receptor Interface} \cap \text{Predicted Receptor Interface}|}
{|\text{Predicted Receptor Interface}|}$} \\

Receptor Recall (RR) &
{\tiny $\displaystyle \frac{|\text{True Receptor Interface} \cap \text{Predicted Receptor Interface}|}
{|\text{True Receptor Interface}|}$} \\

Ligand Precision (LP) &
{\tiny $\displaystyle \frac{|\text{Predicted Ligand Interface} \cap \text{True Receptor Interface}|}
{|\text{Predicted Ligand Interface}|}$} \\

Ligand Recall (LR) &
{\tiny $\displaystyle \frac{|\text{True Ligand Interface} \cap \text{Predicted Receptor Interface}|}
{|\text{True Ligand Interface}|}$} \\
\bottomrule
\end{tabular}
\caption{Structural Interface Metrics}
\label{tb:interface_metrix}
\end{table}

\subsubsection*{4.3 Performance Analysis Across Functional Uncertainty Regimes}
\label{app:definition}
To assess PSPM behavior under varying degrees of functional uncertainty, we stratify evaluation using $\mathrm{FUS}_T(\tau)$ at three levels: full sequence (no stratification), $\mathrm{FUS}_T(\tau{=}30)$, and $\mathrm{FUS}_T(\tau{=}50)$. Increasing $\tau$ progressively removes regions that are structurally ambiguous, disorder-dominated, or weakly coupled to downstream task objectives. This setup allows us to isolate how model predictions change as functional uncertainty decreases, rather than merely filtering by confidence alone.

\paragraph{Comparison of Structural Prediction}
Across $\mathrm{FUS}_T(\tau)$ regimes, most models exhibit improved interface metrics as functional uncertainty decreases, indicating that uncertainty-dominated regions pose substantial challenges for accurate interface prediction. Boltz and Chai consistently achieve strong RR/RP/LR/LP scores even at low $\tau$, indicating robustness to uncertainty and flexible interface geometry. In contrast, OpenFold and UniFold exhibit marked degradation on unfiltered sequences and recover only at high $\tau$, suggesting heavier reliance on rigid backbone geometry.
These results highlight a critical limitation of confidence-only evaluation: models that appear competitive under aggressive filtering may fail in uncertainty-rich regimes that dominate real biological interactions.

\begin{table}[h!]
\centering
\small
\setlength{\tabcolsep}{4pt}
\begin{tabular}{l|cccc}
\toprule
\textbf{Model} & \textbf{RR} & \textbf{RP} & \textbf{LR} & \textbf{LP} \\
\midrule
AF2       & \res{0.749}{0.0217} & \res{0.7186}{0.0226} & \res{0.7052}{0.0215} & \res{0.7486}{0.0215} \\
Boltz     & \res{0.7648}{0.0155} & \res{0.7666}{0.0153} & \res{0.7703}{0.0155} & \res{0.7624}{0.0140} \\
Chai      & \res{0.7583}{0.0158} & \res{0.7746}{0.0149} & \res{0.7674}{0.0155} & \res{0.7571}{0.0155} \\
OpenFold  & \res{0.6626}{0.0271} & \res{0.6127}{0.0312} & \res{0.6551}{0.0274} & \res{0.6322}{0.0280} \\
Proteinx  & \res{0.7344}{0.0164} & \res{0.7427}{0.0162} & \res{0.7385}{0.0159} & \res{0.731}{0.0159} \\
UniFold   & \res{0.5876}{0.0593} & \res{0.5431}{0.0604} & \res{0.5892}{0.0711} & \res{0.6116}{0.0609} \\
\bottomrule
\end{tabular}
\caption{RR/RP/LR/LP on full sequences (Original), reported as \res{mean}{95\% CI}.}
\label{tab:original_rl_metrics}
\end{table}

\begin{table}[h!]
\centering
\small
\setlength{\tabcolsep}{4pt}
\begin{tabular}{l|cccc}
\toprule
\textbf{Model} & \textbf{RR} & \textbf{RP} & \textbf{LR} & \textbf{LP} \\
\midrule
AF2       & \res{0.7729}{0.0208} & \res{0.7426}{0.0217} & \res{0.7313}{0.0207} & \res{0.7719}{0.0204} \\
Boltz     & \res{0.7876}{0.0147} & \res{0.7899}{0.0146} & \res{0.7934}{0.0148} & \res{0.7863}{0.0142} \\
Chai      & \res{0.7815}{0.015} & \res{0.798}{0.0142} & \res{0.7898}{0.0147} & \res{0.7804}{0.0148} \\
OpenFold  & \res{0.6904}{0.0263} & \res{0.6386}{0.0306} & \res{0.6828}{0.0267} & \res{0.6593}{0.0272} \\
Proteinx  & \res{0.7584}{0.0158} & \res{0.7658}{0.0155} & \res{0.7626}{0.0152} & \res{0.7544}{0.0155} \\
UniFold   & \res{0.6127}{0.0576} & \res{0.5671}{0.059} & \res{0.6172}{0.0707} & \res{0.6384}{0.0595} \\
\bottomrule
\end{tabular}
\caption{RR/RP/LR/LP on structured regions ($\mathrm{FUS}_T(\tau=30)$ ), reported as \res{mean}{95\% CI}.}
\label{tab:pLDDT30_rl_metrics}
\end{table}

\begin{table}[h!]
\centering
\small
\setlength{\tabcolsep}{4pt}
\begin{tabular}{l|cccc}
\toprule
\textbf{Model} & \textbf{RR} & \textbf{RP} & \textbf{LR} & \textbf{LP} \\
\midrule
AF2       & \res{0.7959}{0.0197} & \res{0.7663}{0.0207} & \res{0.7569}{0.0198} & \res{0.7946}{0.0193} \\
Boltz     & \res{0.8098}{0.0139} & \res{0.8121}{0.0138} & \res{0.8152}{0.014} & \res{0.8093}{0.0134} \\
Chai      & \res{0.8034}{0.0142} & \res{0.8203}{0.0134} & \res{0.8117}{0.0139} & \res{0.8029}{0.014} \\
OpenFold  & \res{0.7182}{0.0254} & \res{0.6644}{0.0297} & \res{0.7111}{0.0258} & \res{0.6861}{0.0263} \\
Proteinx  & \res{0.7811}{0.015} & \res{0.7884}{0.0147} & \res{0.7857}{0.0144} & \res{0.7775}{0.0144} \\
UniFold   & \res{0.6386}{0.0557} & \res{0.5923}{0.0576} & \res{0.6426}{0.0689} & \res{0.6655}{0.0578} \\
\bottomrule
\end{tabular}
\caption{RR/RP/LR/LP on highly structured regions ($\mathrm{FUS}_T(\tau=50)$ ), reported as \res{mean}{95\% CI}.}
\label{tab:pLDDT50_rl_metrics}
\end{table}

\paragraph{MSE of Training Data on Drug Discovery}
Table~\ref{tab:mse_metrics} reports MSE under increasing $\mathrm{FUS}_T(\tau)$ thresholds. Most models exhibit limited variation across $\tau$, suggesting that excluding high-uncertainty regions does not trivially improve regression performance. Importantly, This apparent stability masks substantial differences in how models handle uncertainty across architectures: AF3, Boltz, and Chai maintain low error even on unfiltered sequences, whereas RoseTTAFold and UniFold show persistently high MSE regardless of $\tau$.
These findings indicate that robustness to functional uncertainty—rather than performance on selectively filtered regions—is a key determinant of practical utility in drug discovery settings.

\begin{table}[h!]
\centering
\small
\setlength{\tabcolsep}{4pt}
\begin{tabular}{l|ccc}
\toprule
\textbf{Model} & \textbf{Original)} & \textbf{$\mathrm{FUS}_T(\tau=30)$ } & \textbf{$\mathrm{FUS}_T(\tau=50)$ } \\
\midrule
AF2       & 0.0324 & 0.0322 & 0.0322 \\
AF3       & 0.0049 & 0.0047 & 0.004 \\
Boltz     & 0.0049 & 0.0043 & 0.0042 \\
Chai      & 0.0063 & 0.006 & 0.0063 \\
DeepFold  & 0.0288 & 0.0282 & 0.0281 \\
ESMFold  & 0.0485 & 0.048 & 0.0485 \\
OmegaFold  & 0.0225 & 0.0225 & 0.0219 \\
OpenFold  & 0.0257 & 0.0249 & 0.0251 \\
Proteinx  & 0.0168 & 0.0163 & 0.0159 \\
RoseTTAFold   & 0.068 & 0.0678 & 0.0674 \\
UniFold   & 0.0529 & 0.0528 & 0.0521 \\
\bottomrule
\end{tabular}
\caption{Performance comparison of MSE on drug discovery prediction of training data across PSPMs under
different $\mathrm{FUS}_T(\tau)$  thresholds.}
\label{tab:mse_metrics}
\end{table}

\subsubsection*{4.4 Suitability of $\mathrm{FUS}_T(\tau)$  Stratification}
\label{app:plddt}
Our adoption of $\mathrm{FUS}_T(\tau)$ is motivated by both empirical analysis (Figure~\ref{fig:case_fus_disprot}) and prior evidence that disorder-aware evaluation is essential for functional tasks. While pLDDT, actifpTM, and PAE capture complementary notions of structural confidence, none of these signals alone align reliably with downstream task performance.
We therefore evaluated alternative stratification strategies using actifpTM, PAE, and their combinations (Tables~\ref{tab:ppi-actifptm-pae}–\ref{tab:drug-combo}). Across both PPI and drug discovery tasks, all alternatives underperform $\mathrm{FUS}_T(\tau)$, despite being more restrictive. This demonstrates that $\mathrm{FUS}_T(\tau)$ captures a task-relevant uncertainty signal that is more aligned with downstream functional performance than existing confidence-based filters.
As a result, $\mathrm{FUS}_T(\tau)$ provides a principled and empirically justified stratification mechanism that is better aligned with biological function than existing confidence-based filters.

\begin{table*}[h!]
\centering
\scriptsize
\setlength{\tabcolsep}{4pt}
\renewcommand{\arraystretch}{1.1}
\caption{PPI performance under different metric thresholds: actifpTM and PAE.}
\label{tab:ppi-actifptm-pae}
\begin{tabular}{lcccccccccccccccc}
\toprule
\multirow{2}{*}{\textbf{PSPM}} & \multicolumn{4}{c}{\textbf{actifpTM $\geq$ 0.7}} & \multicolumn{4}{c}{\textbf{actifpTM $\geq$ 0.9}} & \multicolumn{4}{c}{\textbf{PAE $<$ 15\AA}} & \multicolumn{4}{c}{\textbf{PAE $<$ 10\AA}} \\
\cmidrule(lr){2-5} \cmidrule(lr){6-9} \cmidrule(lr){10-13} \cmidrule(lr){14-17}
 & Acc & Prec & Rec & F1 & Acc & Prec & Rec & F1 & Acc & Prec & Rec & F1 & Acc & Prec & Rec & F1 \\
\midrule
AF2       & 0.79 & 0.78 & 0.80 & 0.79 & 0.80 & 0.79 & 0.81 & 0.80 & 0.79 & 0.78 & 0.80 & 0.79 & 0.80 & 0.78 & 0.80 & 0.80 \\
AF3       & 0.89 & 0.88 & 0.89 & 0.90 & 0.90 & 0.89 & 0.90 & 0.91 & 0.89 & 0.88 & 0.89 & 0.90 & 0.90 & 0.88 & 0.90 & 0.90 \\
Boltz     & 0.85 & 0.84 & 0.85 & 0.85 & 0.86 & 0.85 & 0.86 & 0.86 & 0.85 & 0.84 & 0.85 & 0.85 & 0.86 & 0.84 & 0.86 & 0.86 \\
Chai      & 0.85 & 0.84 & 0.86 & 0.85 & 0.86 & 0.85 & 0.87 & 0.86 & 0.85 & 0.84 & 0.86 & 0.85 & 0.86 & 0.84 & 0.86 & 0.86 \\
OpenFold  & 0.63 & 0.61 & 0.63 & 0.62 & 0.64 & 0.62 & 0.64 & 0.63 & 0.63 & 0.61 & 0.63 & 0.62 & 0.64 & 0.62 & 0.64 & 0.62 \\
Proteinx  & 0.81 & 0.81 & 0.81 & 0.81 & 0.82 & 0.82 & 0.82 & 0.82 & 0.81 & 0.81 & 0.81 & 0.81 & 0.82 & 0.82 & 0.82 & 0.82 \\
UniFold   & 0.56 & 0.38 & 0.66 & 0.48 & 0.57 & 0.39 & 0.67 & 0.49 & 0.56 & 0.38 & 0.66 & 0.48 & 0.56 & 0.38 & 0.66 & 0.48 \\
\bottomrule
\end{tabular}
\end{table*}

\begin{table*}[h!]
\centering
\scriptsize
\setlength{\tabcolsep}{4pt}
\renewcommand{\arraystretch}{1.1}
\caption{PPI performance under combined structural filters: $\mathrm{FUS}_T(\tau=50)$ , actifpTM $\geq$ 0.9, and PAE$<$10Å.}
\label{tab:ppi-combo}
\begin{tabular}{lcccccccccccccccc}
\toprule
\multirow{2}{*}{\textbf{PSPM}} & \multicolumn{4}{c}{\textbf{$\mathrm{FUS}_T(\tau)$  \& actifpTM}} & \multicolumn{4}{c}{\textbf{actifpTM \& PAE}} & \multicolumn{4}{c}{\textbf{$\mathrm{FUS}_T(\tau)$  \& PAE}} & \multicolumn{4}{c}{\textbf{$\mathrm{FUS}_T(\tau)$  \& PAE \& actifpTM}} \\
\cmidrule(lr){2-5} \cmidrule(lr){6-9} \cmidrule(lr){10-13} \cmidrule(lr){14-17}
 & Acc & Prec & Rec & F1 & Acc & Prec & Rec & F1 & Acc & Prec & Rec & F1 & Acc & Prec & Rec & F1 \\
\midrule
AF2       & 0.80 & 0.79 & 0.81 & 0.80 & 0.80 & 0.78 & 0.80 & 0.80 & 0.80 & 0.78 & 0.80 & 0.80 & 0.80 & 0.78 & 0.80 & 0.80 \\
AF3       & 0.90 & 0.89 & 0.90 & 0.91 & 0.90 & 0.88 & 0.90 & 0.90 & 0.90 & 0.88 & 0.90 & 0.90 & 0.90 & 0.88 & 0.90 & 0.90 \\
Boltz     & 0.86 & 0.85 & 0.86 & 0.86 & 0.86 & 0.84 & 0.86 & 0.86 & 0.86 & 0.84 & 0.86 & 0.86 & 0.86 & 0.84 & 0.86 & 0.86 \\
Chai      & 0.86 & 0.85 & 0.87 & 0.86 & 0.86 & 0.84 & 0.86 & 0.86 & 0.86 & 0.84 & 0.86 & 0.86 & 0.86 & 0.84 & 0.86 & 0.86 \\
OpenFold  & 0.64 & 0.62 & 0.64 & 0.63 & 0.64 & 0.62 & 0.64 & 0.62 & 0.64 & 0.62 & 0.64 & 0.62 & 0.64 & 0.62 & 0.64 & 0.62 \\
Proteinx  & 0.82 & 0.82 & 0.82 & 0.82 & 0.82 & 0.82 & 0.82 & 0.82 & 0.82 & 0.82 & 0.82 & 0.82 & 0.82 & 0.82 & 0.82 & 0.82 \\
UniFold   & 0.57 & 0.39 & 0.67 & 0.49 & 0.56 & 0.38 & 0.66 & 0.48 & 0.56 & 0.38 & 0.66 & 0.48 & 0.56 & 0.38 & 0.66 & 0.48 \\
\bottomrule
\end{tabular}
\end{table*}

\begin{table*}[h!]
\centering
\scriptsize
\setlength{\tabcolsep}{4pt}
\renewcommand{\arraystretch}{1.1}
\caption{Drug discovery performance under different metric thresholds: actifpTM and PAE.}
\label{tab:drug-actifptm-pae}
\begin{tabular}{lcccccccc}
\toprule
\multirow{2}{*}{\textbf{PSPM}} & \multicolumn{2}{c}{\textbf{actifpTM $\geq$ 0.7}} & \multicolumn{2}{c}{\textbf{actifpTM $\geq$ 0.9}} & \multicolumn{2}{c}{\textbf{PAE $<$ 15\AA}} & \multicolumn{2}{c}{\textbf{PAE $<$ 10\AA}} \\
\cmidrule(lr){2-3} \cmidrule(lr){4-5} \cmidrule(lr){6-7} \cmidrule(lr){8-9}
 & MAE & $R$ & MAE & $R$ & MAE & $R$ & MAE & $R$ \\
\midrule
AF2         & 0.173 & 0.979 & 0.170 & 0.982 & 0.168 & 0.980 & 0.163 & 0.983 \\
AF3         & 0.062 & 0.994 & 0.059 & 0.996 & 0.061 & 0.991 & 0.058 & 0.993 \\
Boltz       & 0.087 & 0.990 & 0.085 & 0.991 & 0.099 & 0.987 & 0.089 & 0.989 \\
Chai        & 0.102 & 0.990 & 0.101 & 0.991 & 0.109 & 0.988 & 0.105 & 0.991 \\
OmegaFold   & 0.163 & 0.982 & 0.165 & 0.981 & 0.148 & 0.981 & 0.146 & 0.984 \\
OpenFold    & 0.162 & 0.983 & 0.167 & 0.982 & 0.159 & 0.980 & 0.165 & 0.981 \\
Proteinx    & 0.163 & 0.984 & 0.168 & 0.983 & 0.162 & 0.982 & 0.167 & 0.981 \\
DeepFold    & 0.160 & 0.985 & 0.166 & 0.984 & 0.164 & 0.983 & 0.165 & 0.984 \\
ESMFold     & 0.078 & 0.992 & 0.079 & 0.990 & 0.080 & 0.991 & 0.079 & 0.992 \\
RoseTTAFold & 0.200 & 0.974 & 0.207 & 0.973 & 0.198 & 0.972 & 0.201 & 0.974 \\
UniFold     & 0.190 & 0.978 & 0.198 & 0.977 & 0.192 & 0.976 & 0.197 & 0.975 \\
\bottomrule
\end{tabular}
\end{table*}

\begin{table*}[h!]
\centering
\scriptsize
\setlength{\tabcolsep}{4pt}
\renewcommand{\arraystretch}{1.1}
\caption{Drug discovery performance under combined structural filters: $\mathrm{FUS}_T(\tau=50)$ , actifpTM $\geq$ 0.9, and PAE$<$10Å.}
\label{tab:drug-combo}
\begin{tabular}{lcccccccc}
\toprule
\multirow{2}{*}{\textbf{PSPM}} & \multicolumn{2}{c}{\textbf{$\mathrm{FUS}_T(\tau)$  \& actifpTM}} & \multicolumn{2}{c}{\textbf{actifpTM \& PAE}} & \multicolumn{2}{c}{\textbf{$\mathrm{FUS}_T(\tau)$  \& PAE}} & \multicolumn{2}{c}{\textbf{$\mathrm{FUS}_T(\tau)$  \& actifpTM \& PAE}} \\
\cmidrule(lr){2-3} \cmidrule(lr){4-5} \cmidrule(lr){6-7} \cmidrule(lr){8-9}
 & MAE & $R$ & MAE & $R$ & MAE & $R$ & MAE & $R$ \\
\midrule
AF2         & 0.173 & 0.977 & 0.165 & 0.979 & 0.175 & 0.984 & 0.164 & 0.984 \\
AF3         & 0.058 & 0.993 & 0.067 & 0.994 & 0.067 & 0.997 & 0.050 & 0.994 \\
Boltz       & 0.087 & 0.993 & 0.090 & 0.989 & 0.085 & 0.990 & 0.087 & 0.993 \\
Chai        & 0.106 & 0.992 & 0.104 & 0.992 & 0.117 & 0.990 & 0.104 & 0.990 \\
OmegaFold   & 0.146 & 0.980 & 0.162 & 0.983 & 0.157 & 0.982 & 0.169 & 0.981 \\
OpenFold    & 0.159 & 0.982 & 0.165 & 0.980 & 0.160 & 0.981 & 0.164 & 0.981 \\
Proteinx    & 0.160 & 0.983 & 0.166 & 0.981 & 0.162 & 0.982 & 0.168 & 0.980 \\
DeepFold    & 0.159 & 0.984 & 0.161 & 0.983 & 0.164 & 0.982 & 0.165 & 0.983 \\
ESMFold     & 0.072 & 0.995 & 0.074 & 0.993 & 0.073 & 0.994 & 0.075 & 0.993 \\
RoseTTAFold & 0.195 & 0.973 & 0.194 & 0.972 & 0.196 & 0.974 & 0.197 & 0.973 \\
UniFold     & 0.180 & 0.976 & 0.188 & 0.975 & 0.186 & 0.976 & 0.190 & 0.974 \\
\bottomrule
\end{tabular}
\end{table*}

\subsubsection*{4.5 Compute Infrastructure and Resource Utilization}

All experiments in our benchmark were conducted using NVIDIA A100 GPUs. We adopt the default configuration settings provided by each model to ensure consistency and ease of replication. The total computation time across all tasks amounts to approximately 2,000 GPU hours. Detailed instructions, including execution scripts, are provided in our repository to support reproducibility on comparable hardware.

\subsection*{5. Visual Analytics Portal for Structure and Task Evaluation}
We provide detailed instructions for using the portal to obtain desired protein structures and compare performance across downstream tasks. The portal consists of three primary components: (1) User Input Panel – Includes a navigation panel, a protein ID input field, and a PSPM (Protein Structure Prediction Model) selection interface. (2) Visualization Panel – Contains a Root Mean Square Deviation (RMSD)-based alignment heatmap and a 3D protein structure viewer. (3) Downstream Task Comparison Panel – Displays prediction results from selected PSPM models for relevant biological tasks.

\begin{figure}[h!]
    \centering
    \includegraphics[width=\linewidth]{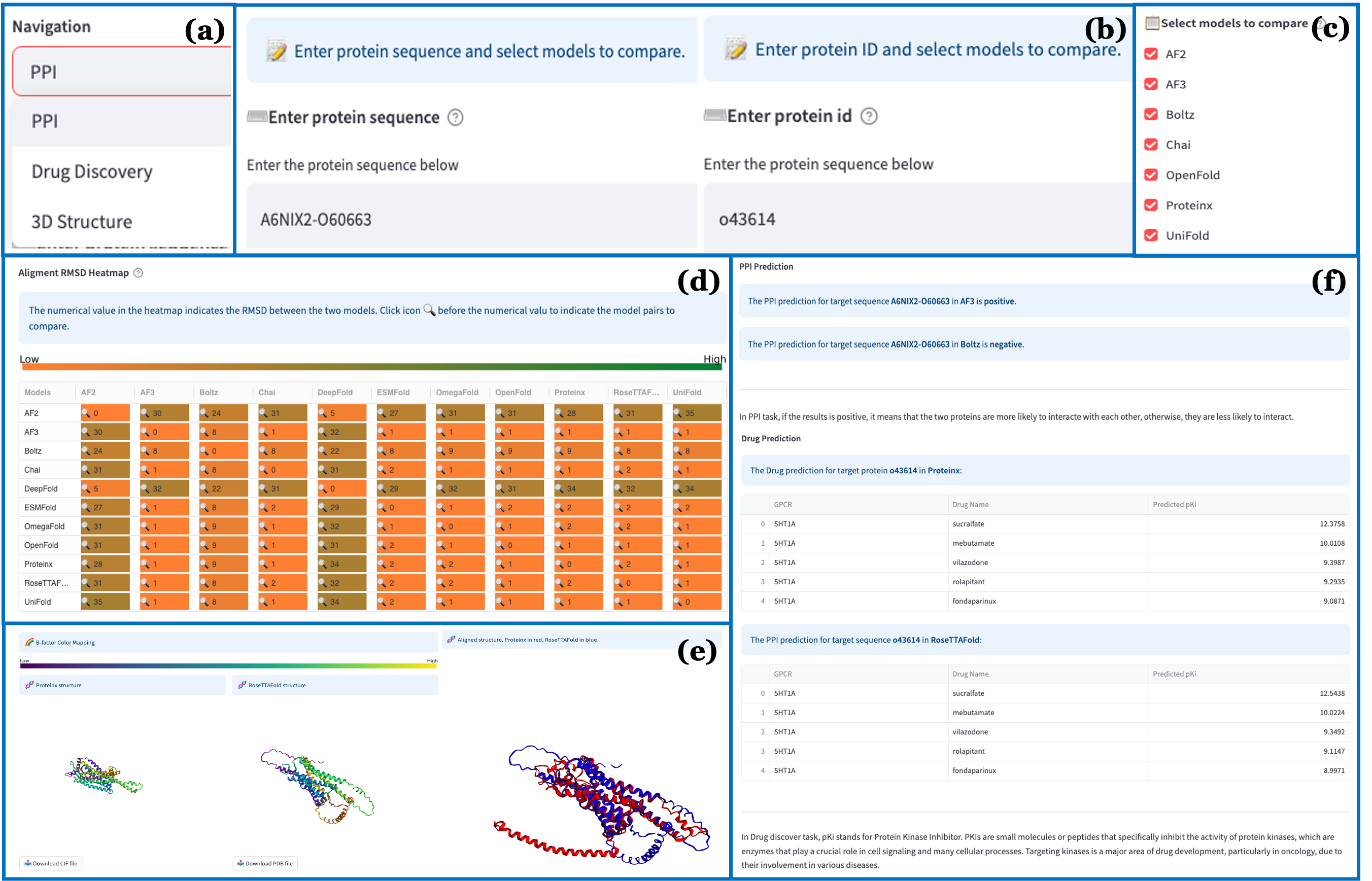}
    \caption{Portal Overview. (a) Navigation for task type. (b) Protein input. (c) Model selection. (d) Structural alignment heatmap. (e) 3D visualization and residue-level uncertainty. (f) Downstream task predictions.}
    \label{fig: UI_app}
\end{figure}

A screenshot of the portal interface is shown in Figure~\ref{fig: UI_app}. Users can follow the steps below to interact with the portal:

1. \textbf{Select a Downstream Task}
Users begin by selecting the downstream task of interest. The portal currently supports three options: PPI (Protein-Protein Interaction), Drug Discovery, 3D Structure Access (providing access to a database of 3D protein structures and intrinsically disordered regions, or IDRs)

2. \textbf{Input Protein ID and Select PSPMs}
Once a task is selected, the protein ID input panel (Figure~\ref{fig: UI_app}.b) and the PSPM selection panel (Figure~\ref{fig: UI_app}.e) become available. Users must enter a valid protein ID according to the input format hint and choose at least two PSPMs for comparison.

3. \textbf{Visualize RMSD Alignment and Protein Structures}
Upon submission, the portal computes an RMSD alignment heatmap, which is displayed in the heatmap panel (Figure~\ref{fig: UI_app}.d). Users can explore structure comparisons by selecting a pair of PSPMs using the magnifier icon. This action reveals the corresponding 3D protein structures in the visualization panel (Figure~\ref{fig: UI_app}.e).

4. \textbf{View Downstream Task Results}
Below the visualization panel, prediction results related to the selected downstream task are displayed:

\indent $\bullet$ For PPI, predictions are shown as binary outcomes (positive or negative).

\indent $\bullet$ For Drug Discovery, results are presented in a tabular format listing the top-ranked drug candidates.

\end{document}